\documentclass[aps,pre,twocolumn,superscriptaddress,showpacs]{revtex4-1}
\usepackage{amsmath}
\usepackage{amssymb}
\usepackage{tikz}
\usepackage{graphicx}
\usepackage{dcolumn}
\usepackage{bm}
\usepackage{epsfig}
\usepackage{psfrag}
\def\II{\hbox{$1\hskip -1.2pt\vrule depth 0pt height 1.6ex width 0.7pt\vrule depth 0pt height 0.3pt width 0.12em$}}
\newcommand{\reffig}[1]{\mbox{Fig.~\ref{#1}}}
\newcommand{\refeq}[1]{\mbox{Eq.~(\ref{#1})}}
\newcommand{\refsec}[1]{\mbox{Sec.~\ref{#1}}}

\newcommand{\be}{\begin{equation}}
\newcommand{\ee}{\end{equation}}
\newcommand{\ba}{\begin{eqnarray}}
\newcommand{\ea}{\end{eqnarray}}

\newcommand{\T}{${\mathcal T}\,$}

\def\II{\hbox{$1\hskip -1.2pt\vrule depth 0pt height 1.6ex width 0.7pt\vrule depth 0pt height 0.3pt width 0.12em$}}

\begin{document}

\title{\bf Missing-level statistics in classically chaotic quantum systems with symplectic symmetry}
\author{Jiongning Che}
\affiliation{Lanzhou Center for Theoretical Physics and the Gansu Provincial Key
Laboratory of Theoretical Physics, Lanzhou University, Lanzhou University, Lanzhou, Gansu 730000, China}
\author{Junjie Lu}
\affiliation{Lanzhou Center for Theoretical Physics and the Gansu Provincial Key
Laboratory of Theoretical Physics, Lanzhou University, Lanzhou University, Lanzhou, Gansu 730000, China}
\affiliation{Institut de Physique de Nice, CNRS UMR 7010, Universit\'e C${\hat o}$te d'Azur, 06108 Nice, France}
\author{Xiaodong Zhang}
\affiliation{Lanzhou Center for Theoretical Physics and the Gansu Provincial Key
Laboratory of Theoretical Physics, Lanzhou University, Lanzhou University, Lanzhou, Gansu 730000, China}
\author{Barbara Dietz}
\email{dietz@lzu.edu.cn}
\affiliation{Lanzhou Center for Theoretical Physics and the Gansu Provincial Key
Laboratory of Theoretical Physics, Lanzhou University, Lanzhou University, Lanzhou, Gansu 730000, China}
\author{Guozhi Chai}
\affiliation{School of Physical Science and Technology, and Key Laboratory for Magnetism and Magnetic Materials of MOE, Lanzhou University, Lanzhou, Gansu 730000, China}

\date{\today}

\bigskip

\begin{abstract}
	We present experimental and theoretical results for the fluctuation properties in the incomplete spectra of quantum systems with symplectic symmetry and a chaotic dynamics in the classical limit. To obtain theoretical predictions, we extend the random-matrix theory (RMT) approach introduced in [O. Bohigas and M. P. Pato, Phys. Rev. E {\bf 74}, 036212 (2006)] for incomplete spectra of quantum systems with orthogonal symmetry. We validate these RMT predictions by randomly extracting a fraction of levels from complete sequences obtained numerically for quantum graphs and experimentally for microwave networks with symplectic symmetry and then apply them to incomplete experimental spectra to demonstrate their applicability. Independently of their symmetry class quantum graphs exhibit nongeneric features which originate from nonuniversal contributions. Part of the associated eigenfrequencies can be identified in the level dynamics of parameter-dependent quantum graphs and extracted, thereby yielding spectra with systematically missing eigenfrequencies. We demonstrate that, even though the RMT approach relies on the assumption that levels are missing at random, it is possible to determine the fraction of missing levels and assign the appropriate symmetry class by comparison of their fluctuation properties with the RMT predictions.  
\end{abstract}

\bigskip
\maketitle

\section{Introduction\label{Intro}}
The manifestation of characteristics of a classical dynamics in the spectral properties of the corresponding quantum system, like nuclei, atoms, molecules, quantum wires and dots or other complex systems~\cite{Brody1981,Zimmermann1988,Guhr1989,Weidenmueller2009,Gomez2011,Frisch2014,Mur2015,Dietz2017,Naubereit2018} is well understood by now. It has, for instance, been established that the spectral properties of generic quantum systems with chaotic classical counterpart are universal. According to the Bohigas-Giannoni-Schmit (BGS) conjecture~\cite{Berry1977a,Berry1979,Casati1980,Bohigas1984} they coincide with those of random matrices from the Gaussian ensembles of corresponding universality class~\cite{Mehta1990}, as proven rigorously based on a semiclassical approach~\cite{Heusler2007}. Berry and Tabor showed in~\cite{Berry1977a} that for typical systems with integrable classical dynamics (see Ref.~\cite{Robnik1998} for a detailed specification of 'typical') they agree well with those of Poissonian random numbers, explicitely excluding the harmonic oscillator, which is a paradigm example for an 'untypical' system~\cite{Drod1991}. Numerous studies with focus on problems from the field of quantum chaos have been performed both theoretically and experimentally with microwave billiards~\cite{Sridhar1991,Graef1992,Stein1992,So1995,Deus1995,Stoeckmann2001,Dietz2015a} and microwave networks~\cite{Hul2004,Lawniczak2010} as model systems. In the experiments, the analogy between the Helmholtz equation for flat microwave cavities and networks of coaxial cables with the Schr\"odinger equation for quantum billiards~\cite{LesHouches1989,Stoeckmann2001,Haake2018,Texier2001} and quantum graphs~\cite{Kottos1997,Kottos1999,Pakonski2001}, respectively, is exploited. 

Quantum graphs were originally proposed by Linus Pauling to emulate certain features of organic molecules \cite{Pauling1936} and are used as models for quantum wires~\cite{Sanchez1998}, optical waveguides and mesoscopic quantum systems~\cite{Kowal1990}. They also serve as an ideal testbed for the investigation of universal properties of closed and open quantum systems with chaotic classical dynamics~\cite{Stoeckmann1999,Haake2018}. Indeed, the spectral properties of closed quantum graphs with incommensurable bond lengths were proven rigorously to coincide with those of random matrices from the Gaussian ensemble of the same universality class~\cite{Gnutzmann2004,Pluhar2014}. Furthermore, the semiclassical approximation of their spectral density in terms of classical periodic orbits is exact~\cite{Keating1991,Kottos1999} and the correlation functions of the scattering matrix elements of open quantum graphs coincide with the corresponding random matrix theory (RMT) results~\cite{Verbaarschot1985,Pluhar2013,Pluhar2013a,Pluhar2014,Fyodorov2005} for quantum chaotic scattering systems. From the experimental point of view, their most important property is that quantum graphs belonging to the orthogonal, the unitary and the symplectic universality class can be realized with microwave networks of coaxial cables~\cite{Hul2004,Lawniczak2010,Hul2012,Allgaier2014,Bialous2016,Rehemanjiang2016,Rehemanjiang2018,Martinez2018,Rehemanjiang2020,Lu2020}. In quantum systems with unitary symmetry time-reversal ($\mathcal{T}$) invariance is violated. Systems belonging to the orthogonal or the symplectic universality class preserve $\mathcal{T}$ invariance, where, in the orthogonal case $\mathcal{T}^2=1$ and in the symplectic one $\mathcal{T}^2=-1$, corresponding to integer and half-integer spin systems, respectively~\cite{Scharf1988,Haake2018}. Note, that the eigenvalues of systems with symplectic symmetry exhibit Kramer's degeneracy, so that in numerics and experiment only half of them are found. According to the BGS conjecture the spectral properties of quantum systems with chaotic classical counterpart and orthogonal, unitary or symplectic symmetry  coincide with those of random matrices from the Gaussian orthogonal ensemble (GOE), the Gaussian unitary ensemble (GUE) or the Gaussian symplectic ensemble (GSE), respectively. 

Yet, quantum graphs have one drawback. Namely, they exhibit nongeneric features originating from nonuniversal contributions of eigenstates, which due to backscattering at the vertices terminating them are localized on individual bonds, closed loops or combinations of loops within the quantum graph, that is, on a fraction of it~\cite{Kottos1999,Dietz2017}. These loop states lead to topological resonances~\cite{Gnutzmann2013} in open quantum graphs and to deviations of the spectral properties from RMT predictions in closed ones. Modes which are localized on a small part of the quantum graph are nongeneric because they do not sense the chaoticity of the underlying classical dynamics, which results from the scattering at all vertices. The associated eigenenergies depend on the lengths of the associated bonds, implying that such modes are nonuniversal. Their effect on the spectral properties is, e.g., comparable to that of bouncing-ball orbits in a stadium billiard~\cite{Sieber1993}. It becomes visible in the short- and long range correlations in the eigenvalue spectra for level distances larger than about 2-3 mean spacings and thus does not prevent level repulsion or modify its degree, that is, it does not disguise the characteristics that enable the determination of the universality class of a chaotic quantum graph, e.g., from the degree of level repulsion~\cite{Haake2018}. This property, actually, justifies the exclusion of these eigenstates in the proof of the equivalence of the spectral properties of quantum graphs and of random matrices of the Gaussian ensembles in Refs.~\cite{Gnutzmann2004,Pluhar2013,Pluhar2013a,Pluhar2014}. However, these localized states are unavoidable in microwave networks and quantum graphs, since they comply with the particular boundary conditions obeyed by the microwaves or wave functions at the vertices. In Ref.~\cite{Lu2020} modes localized on individual bonds were identified in the level dynamics of parameter-dependent quantum graphs and microwave networks and extracted. This led to an improvement of the agreement of parametric spectral properties with RMT predictions. Extracting eigenfrequencies from a spectrum yields incomplete spectra, yet the RMT predictions for generic quantum systems with chaotic classical dynamics apply only if the spectrum is complete.

Incomplete spectra are, actually, a general problem one has to cope with in experiments with microwave billiards and microwave networks~\cite{Bialous2016,Bialous2016a,Lawniczak2018} and in nuclear, atomic and molecular systems,~\cite{Liou1972,Zimmermann1988,Enders2000,Enders2004,Molina2007,Frisch2014,Mur2015}. An RMT approach was developed in the context of nuclear physics already with the emergence of the field of quantum chaos~\cite{Porter1965,Brody1981,Agvaanluvsan2003,Agvaanluvsan2003a,Bohigas2004,Bohigas2006}. This problem can be circumvented by restricting to statistical measures which do not rely on completeness. This is possible, e.g., in microwave billiards or microwave networks where scattering matrix elements are available whose fluctuation properties also provide measures for the chaoticity, e.g., in terms of their correlation functions, the distributions of their cross sections~\cite{Dietz2009,Dietz2010,Kumar2013,Kumar2017} or the enhancement factor~\cite{Lawniczak2010,Bialous2019,Bialous2020}. Complete spectra were obtained in experiments with superconducting microwave billiards~\cite{Dietz2015,Dietz2019} by choosing either resonators made from niobium or from lead-covered brass. This, however, is impossible for microwave networks constructed from coaxial cables, since they contain a dielectric medium, which prevents superconductivity, even if the cables are made from niobium instead of copper. In Ref.~\cite{Bohigas2006} statistical measures were derived for the fluctuation properties in incomplete spectra of quantum systems with orthogonal symmetry and a chaotic classical counterpart on the basis or RMT. They were extended to systems with unitary symmetry and tested with microwave networks and billiards in~\cite{Bialous2016,Lawniczak2018}. In the present paper we derive RMT predictions for quantum systems with symplectic symmetry by proceeding as in Ref.~\cite{Bohigas2006} and validate them by randomly extracting levels from complete spectra obtained from numerical calculations for quantum graphs and experiments with microwave networks. Furthermore, we will test their applicability using experimental spectra of microwave networks, for which the identification of all eigenfrequencies was not possible and also to the spectra of GSE graphs after extraction of nonuniversal contributions~\cite{Lu2020}. 

We briefly introduce in~\refsec{Theory} microwave networks and quantum graphs. In~\refsec{Missing} we present the RMT approach developed in Ref.~\cite{Bohigas2006} for the fluctuation properties in the spectra of classically chaotic quantum systems. Then, in~\refsec{Test} we test the RMT predictions both numerically and experimentally for various realizations of GSE graphs and also compare them to those obtained for GUE graphs. Finally, the results are discussed in~\refsec{Concl}.

\section{Microwave networks as a model for GSE quantum graphs\label{Theory}} 
A quantum graph consists of $\mathcal{V}$ vertices $i=1,\dots,\mathcal{V}$ that are connected by $\mathcal{B}$ bonds where the wave function component $\psi_{ij}(x)$ on the bond connecting vertices $i$ and $j$ is a solution of the one-dimensional Schr\"odinger equation
\be
-\frac{{\rm d}^2}{{\rm d}x^2}\psi_{ij}(x)=k^2\psi_{ij}(x), 
\ee
with the boundary condition that $\psi_{ij}$ is continuous and current is conserved at the vertices $i$ and $j$. Imposing these boundary conditions on the wave function components $\psi_{ij}$ yields the quantization condition of the quantum graph, that is, an equation for it's eigenwavenumbers $k_n$~\cite{Kottos1999}. A quantum graph is characterized by the lengths $L_{ij}$ of the bonds and the connectivity matrix $\hat C$ with diagonal elements $C_{ii}=0$ and nonzero off-diagonal elements $C_{ij}=1$ for connected vertices $i$ and $j$. It has been shown in Refs.~\cite{Gnutzmann2004,Pluhar2014} that a quantum graph exhibits spectral properties of a typical quantum system with chaotic classical counterpart, if the bond lengths are incommensurable. 

\begin{figure}[!th]
\includegraphics[width=\linewidth]{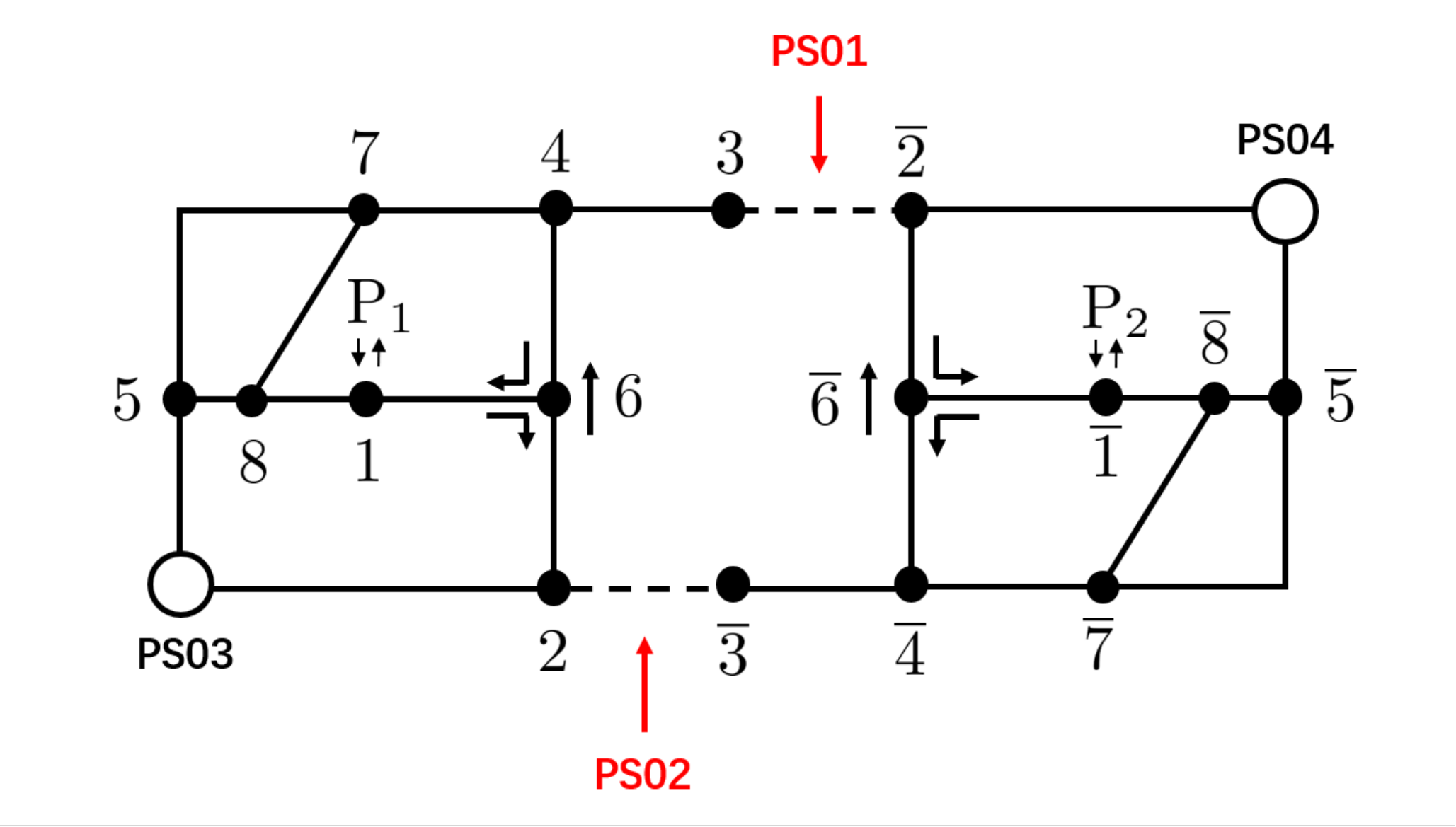}
\includegraphics[width=\linewidth]{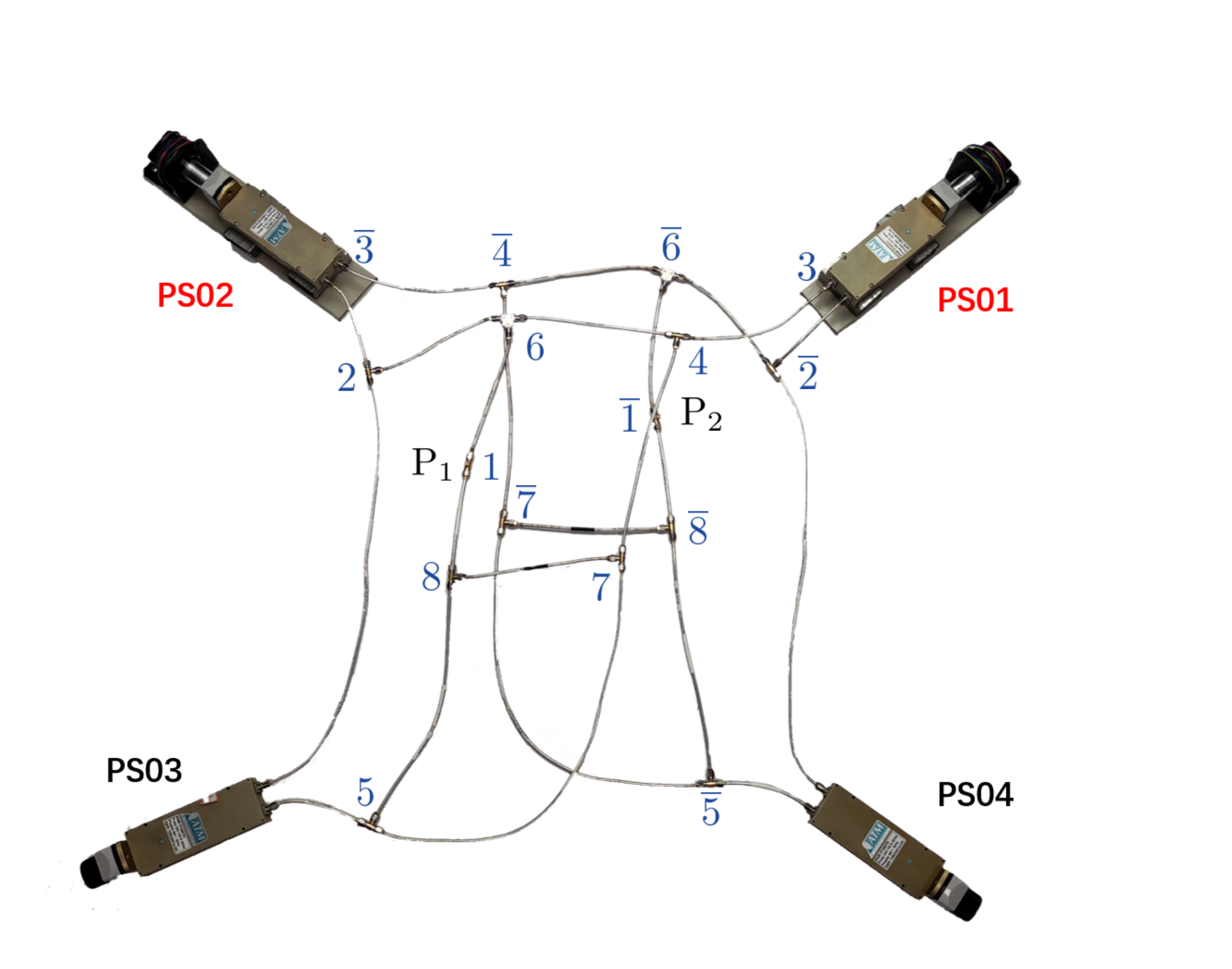}
	\caption{Schematic view (upper panel) of the GSE quantum graph and photograph (lower panel) of the corresponding microwave network. They are constructed from two GUE subgraphs. Here, time-reversal invariance violation is induced by replacing the $T$ joints marked by $6,\, \bar 6$ by $T$-shaped circulators which induce unidirectionality as indicated by the arrows. The two subgraphs are identical except for the orientation of the circulators, where corresponding vertices are marked by $n$ and $\bar n$, with $n=1,\dots,8$. They are connected by two coaxial cables and phasers (marked PS01 and PS02) which induce a relative phase of $\pi$. An ensemble of GSE graphs was realized by increasing the lengths of two corresponding bonds with phasers (marked by PS03 and PS04) by the same amount.} 
\label{Fig1}
\end{figure}
The upper part of~\reffig{Fig1} exhibits one of the GSE graphs, constructed from vertices of valency three, which were used in this paper. The lower part shows the corresponding experimental realization which consists of a network of microwave coaxial cables, whose optical lengths correspond to the bond lengths in the quantum graphs, connected by $T$ joints. The coaxial cables consist of an inner and concentric outer conductor and Teflon with an experimentally determined dielectric constant $\epsilon\simeq 2.06$, which fills the space between them. The analogy to a quantum graph of corresponding geometry holds below the cut-off frequency for the first transverse electric mode~\cite{Jones1964,Savytskyy2001}, where only the fundamental transverse electromagnetic (TEM) mode can propagate in the coaxial cables. Denoting by $U_{ij}(x)$ the difference between the potentials at the conductors' surfaces in the coaxial cable connecting joints $i$ and $j$, by $c$ the velocity of light in vacuum and by $\omega$ the angular frequency, respectively, the associated telegraph equations read 
\begin{equation}
\frac{{\rm d}^2}{{\rm d}x^2}U_{ij}(x)+\frac{\omega^2\epsilon}{c^2}U_{ij}(x)=0.
\label{WE}
\end{equation}
This set of equations is applicable to lossless coaxial cables, that is, for vanishing Ohmic resistance. At the vertices $U_{ij}(x)$ obeys the continuity equation and Neumann boundary conditions implying that current is conserved. Thus, the wave equations $\refeq{WE}$ governing the $U_{ij}(x)$ are mathematically identical to the Schr\"odinger equation of a quantum graph with Neumann boundary condition at the vertices~\cite{Kottos1999,Texier2001} when identifying $\sqrt{\epsilon}\frac{\omega}{c}$ of the microwave network with the wavenumber $k$ of the quantum graph. Hence the eigenfrequencies $\nu_i$ of a microwave network which is constructed from coaxial cables of lengths $\tilde L_{ij}$ with optical length $L_{ij}=\sqrt{\epsilon}\tilde L_{ij}$ yield the eigenwavenumbers $k_i=\frac{2\pi\nu_i}{c}$ of the quantum graph of corresponding connectivity composed of bonds of lengths $L_{ij}$. More details on the experiments are provided in Ref.~\cite{Lu2020}.

The GSE graphs consist of two connected GUE graphs~\cite{Rehemanjiang2016}. Time-reversal invariance violation is induced by $T$-shaped circulators~\cite{Lawniczak2010,Bialous2016} at the vertices 6 and $\bar 6$, which cause unidirectionality of propagation of microwaves through them as indicated by the arrows in~\reffig{Fig1}. The GUE graphs are identical except for the orientations of the circulators, where corresponding vertices are denoted by $n$ and $\bar n$, with $n=1,\dots,8$. They are connected by two bonds of same length. A total relative phase of $\pi$ of the microwaves traveling through the connecting cables was generated by phase shifters (marked PS01 and PS02 in~\reffig{Fig1}) which change the lengths of the coaxial cables by some increment $\Delta \tilde l$ yielding a phase increment
\be
\Delta\varphi=k\Delta\tilde l=\frac{2\pi \nu}{c}\Delta\tilde l.
\label{DeltaL}
\ee
In order to attain an optimum tuning of the relative phase to $\pi$, we measured transmission amplitudes $\vert S_{1\bar 1}(\nu)\vert$ and $\vert S_{\bar 1 1}(\nu)\vert$ between antennas attached at ports P1 and P2 at the $T$-joints marked by $1$ and $\bar1$ in ~\reffig{Fig1} and used the fact that transmission is completely suppressed for a precise relative phase of $\pi$ since then microwaves traveling through the connecting bonds from port P1 to port P2 interfere destructively at port P2 and vice versa.

An ensemble of quantum graphs and corresponding microwave networks was realized by varying the lengths of bonds stepwise by a fixed increment $\Delta l$~\cite{Bialous2016}. This is achieved by introducing two phase shifters, denoted by PS03 and PS04 in~\reffig{Fig1}. In part of the experiments two additional phasers were used and the total length $\mathcal{L}$ of the microwave network was kept fixed. Thus, the average of the integrated spectral density $N(\nu)$, that is, the average number $\bar N(\nu)$ of eigenfrequencies below frequency $\nu$, which is given by Weyl's law,
\begin{equation}
\bar{N} (\nu)=\frac{2\mathcal{L}}{c}\nu,
\label{Weyl}
\end{equation}
didn't change. Accordingly, the lengths of two corresponding bonds were increased in $N_{max}$ steps by an increment $\Delta l$ and decreased for another pair by the same amount~\cite{Kottos1999}. In the experiments described in Ref.~\cite{Lu2020} the lengths of four coaxial cables were changed in $N_{max}=43$ steps of size $\Delta l=0.84$~mm, where the total length of the GSE graph equaled $\mathcal{L}=6.68$~m. 

The eigenfrequencies $\nu_i$ of the microwave networks correspond to the positions of the minima exhibited by the reflection amplitude $\vert S_{11}(\nu)\vert$ when measured as function of the microwave frequency $\nu$. Due to the unavoidable absorption of microwaves in the coaxial cables, these resonances are broadened and thus may overlap depending on the size of absorption, thus turning the identification of eigenfrequencies into a cumbersome, if not impossible task. The problem of absorption has been eliminated in experiments with flat, cylindrical microwave resonators simulating quantum billiards~\cite{Stoeckmann1990,Sridhar1991,Graef1992,So1995} by performing the measurements with superconducting cavities~\cite{Graef1992,Dembowski2002,Dietz2015}. This is not possible with microwave networks, because they contain Teflon. In the experiments described in~\cite{Lu2020} we were able to identify all eigenfrequencies in the frequency range where the analogy to a quantum graph holds by means of the measured level dynamics. Since the average integrated spectral density was kept fixed, missing levels were, e.g., identified at jumps in $\bar N(\nu)$ when comparing the eigenfrequency spectra for two neighboring parameter values $\lambda$. For the experimental investigation of the fluctuation properties in incomplete spectra of quantum systems belonging to the symplectic universality class we used among others the thus obtained complete spectra and randomly extracted up to $30\%$ of the eigenfrequencies.

Furthermore, we used the eigenvalues which were obtained from numerical simulations of parametric quantum graphs in Ref.~\cite{Lu2020}. Here, instead of employing the vertex secular equation deduced from the quantization condition for closed quantum graphs~\cite{Kottos1999} we applied the scattering formalism for open quantum graphs which, actually, is more appropriate for the description of the experimental situation. Namely, for the measurement of reflection and transmission scattering amplitudes the microwave networks are slightly opened through the antennas. Within this scattering approach the eigenvalues of the corresponding closed quantum graph with $\mathcal{B}$ bonds correspond to the solutions of the secular equation for the $2\mathcal{B}\times 2\mathcal{B}$-dimensional bond scattering matrix $\hat S_\mathcal{B}(k;\{\Phi_{ij}\})$~\cite{Texier2001,Kottos2003},
\be
\zeta_\mathcal{B}(k)=\det\left[\II-\hat S_\mathcal{B}(k;\{\Phi_{ij}\})\right]=0,
\label{zeta}
\ee
where
\be
\hat S_\mathcal{B}(k;\{\Phi_{ij}\})=\hat D(k;\{\Phi_{ij}\})\hat T
\ee
in the $2\mathcal{B}$ space of directed bonds, and
\ba 
\hat D_{ij,nm}&=&\delta_{i,n}\delta_{j,m}e^{ikL_{ij}+\Phi_{ij}},\\
\hat T_{ji,nm}&=&\delta_{n,i}C_{j,i}C_{n,m}\hat\sigma^{(i)}_{ji,nm}.
\ea
The relative phase of $\pi$ is accounted for in the phases $\Phi_{ij}$ and the directionality at the circulators is incorporated by appropriately choosing the vertex scattering matrix $\hat\sigma^{(i)}$, which enters the transition matrix $\hat T_{ji,im}$ from vertex $m$ to vertex $j$ via vertex $i$~\cite{Lu2020}.

\section{RMT approach for the spectral properties of incomplete spectra\label{Missing}} 
Before comparing the spectral properties of the quantum graphs with random matrix theory (RMT) predictions for universal quantum systems with chaotic classical counterpart, their system specific properties need to be eliminated. This is done by unfolding their eigenvalues such that their spectral density, that is, their mean spacing is uniform, and rescaling them to mean spacing unity. The mean spectral density of quantum graphs and microwave networks 
\begin{equation}
\bar\rho (\nu)=\frac{2\mathcal{L}}{c},
\label{Weyl2}
\end{equation}
is frequency independent. This implies, that unfolded eigenvalues $\epsilon_i$ of mean spacing unity are attained from the ordered eigenfrequencies $\nu_i$, with their size increasing with the index $i$, by multiplying them with a constant factor, $\epsilon_i=2\nu_i\mathcal{L}/c$.

In order to obtain information on short-range correlations in the eigenvalue spectra of the quantum graphs and microwave networks we analyzed the nearest-neighbor spacing distribution $P(s)$ of adjacent spacings $s_i=\epsilon_{i+1}-\epsilon_i$ and its cumulant $I(s)=\int_0^s{\rm d}s^\prime P(s^\prime)$, which has the advantage that it does not depend on the binning size of the histograms yielding $P(s)$. Furthermore, we considered the variance 
\be
\Sigma^2(L)=\left\langle\left( N(L)-\langle N(L)\rangle\right)^2\right\rangle 
\ee
of the number of unfolded eigenvalues $N(L)$ in an interval of length $L$ where $\langle N(L)\rangle =L$, and the rigidity
\be
\Delta_3(L)=\left\langle \min_{a,b}\int_{e-L/2}^{e+L/2}de\left[N(e)-a-be\right]^2\right\rangle
\ee
which provides information on the stiffness of a spectrum. Here, $\langle\cdot\rangle$ denotes the average over an ensemble of random matrix or quantum graph realizations. These statistical measures provide information on long-range spectral fluctuations, a further one being the power spectrum of a sequence of $N$ levels~\cite{Relano2002,Faleiro2004,Molina2007,Riser2017}. It is given in terms of the Fourier transform of the deviation of the $q$th nearest-neighbor spacing from its mean value $q$, $\delta_q=\epsilon_{q+1}-\epsilon_1-q$, from $q$ to $\tau$
\begin{equation}
\label{PowerS}
S(\tau)=\left\langle\left\vert\frac{1}{\sqrt{N}}\sum_{q=0}^{N-1} \delta_q\exp\left(-\frac{2\pi i\tau q}{N}\right)\right\vert^2\right\rangle .
\end{equation}
It was demonstrated in~\cite{Bialous2016,Dietz2017} that for the GOE and GUE $\Sigma^2(L)$ and $S(\tau)$ are particularly sensitive to missing levels. The power spectrum $S(\tau)$ only depends on the ratio $\tilde\tau =\tau/N$ and exhibits for $\tilde\tau\ll 1$ a power law dependence $\langle\tilde S(\tilde\tau)\rangle\propto (\tilde\tau)^{-\alpha}$~\cite{Relano2002,Faleiro2004}, where for regular systems $\alpha =2$. For chaotic ones $\alpha =1$ independently of whether \T invariance is preserved or not~\cite{Gomez2005,Salasnich2005,Santhanam2005,Relano2008,Faleiro2006}, that is, it does not depend on the underlying universality class. 

The objective of the present paper is the experimental and theoretical investigation of the fluctuation properties in incomplete spectra of GSE quantum graphs. For the derivation of RMT predictions for statistical measures of the spectral properties of such systems we followed the procedure outlined in Ref.~\cite{Bohigas2006}. The number of missing levels is characterized by the fraction $\Phi$ of eigenfrequencies that could be identified. In the case of quantum graphs the expected number of eigenfrequencies is obtained from Weyl's law~\refeq{Weyl}. The procedure is based on the assumption that levels are missing randomly, that is, are extracted randomly from the complete sequence. This implies that the probability to observe a level is $\Phi$, and generally, the joint probability distribution $R_n(E_1,\cdots,E_n){\rm d}E_1\dots{\rm d}E_n$ of finding levels at $E_1,\dots,E_n$ is just reduced by a factor $\Phi^n$, 
\be
\label{rn}
r_n(E_1,\cdots,E_n)=\Phi^nR_n(E_1,\cdots,E_n)
\ee 
yielding for the spectral density~\refeq{Weyl2} $r_1(\nu)=\Phi\bar\rho(\nu)$ so that the mean spacing increases by a factor $\Phi^{-1}$. The nearest-neighbor spacing distribution is expressed in terms of the $(n+1)$st nearest-neighbor spacing distribution $P(n,s)$, with $P(0,s)=P(s)$, of the corresponding complete spectrum,
\ba
\label{abst}
&&p(s)= \sum_{n=0}^{M}(1-\Phi)^nP\left(n;\frac{s}{\Phi}\right)\\
&&\simeq\sum_{n=0}^{K-1}(1-\Phi)^nP\left(n;\frac{s}{\Phi}\right)\label{ps1}\\
&&+\sum_{n=K}^{M}(1-\Phi)^n\frac{1}{\sqrt{2\pi V^2(n)}}\exp\left(-\frac{1}{2V^2(n)}\left[\frac{s}{\Phi}-n-1\right]^2\right),\nonumber
\ea
where
\be
V^2(n)\simeq \Sigma^2(L=n)-\frac{1}{6}.
\ee
Equation (\ref{abst}) is derived from the relation
\be
P(n;s)=\frac{{\rm d}^2}{{\rm d}s^2}\sum_{l=0}^n(n-l+1)E(l;s)
\ee
where $E(l;s)$ is the probability that an interval of length $s$ contains $l$ levels. Using that for randomly missing levels the probability to miss a level is $1-\Phi$ yields for the incomplete sequence 
\be
e(n;s)=\sum_{l=n}^\infty\begin{pmatrix} l\\ n\end{pmatrix}\Phi^n(1-\Phi)^{l-n}E\left(l;\frac{s}{\Phi}\right) 
\ee
and thus for $n=0$~\refeq{abst}.
	
\begin{figure}[!th]
\includegraphics[width=0.9\linewidth]{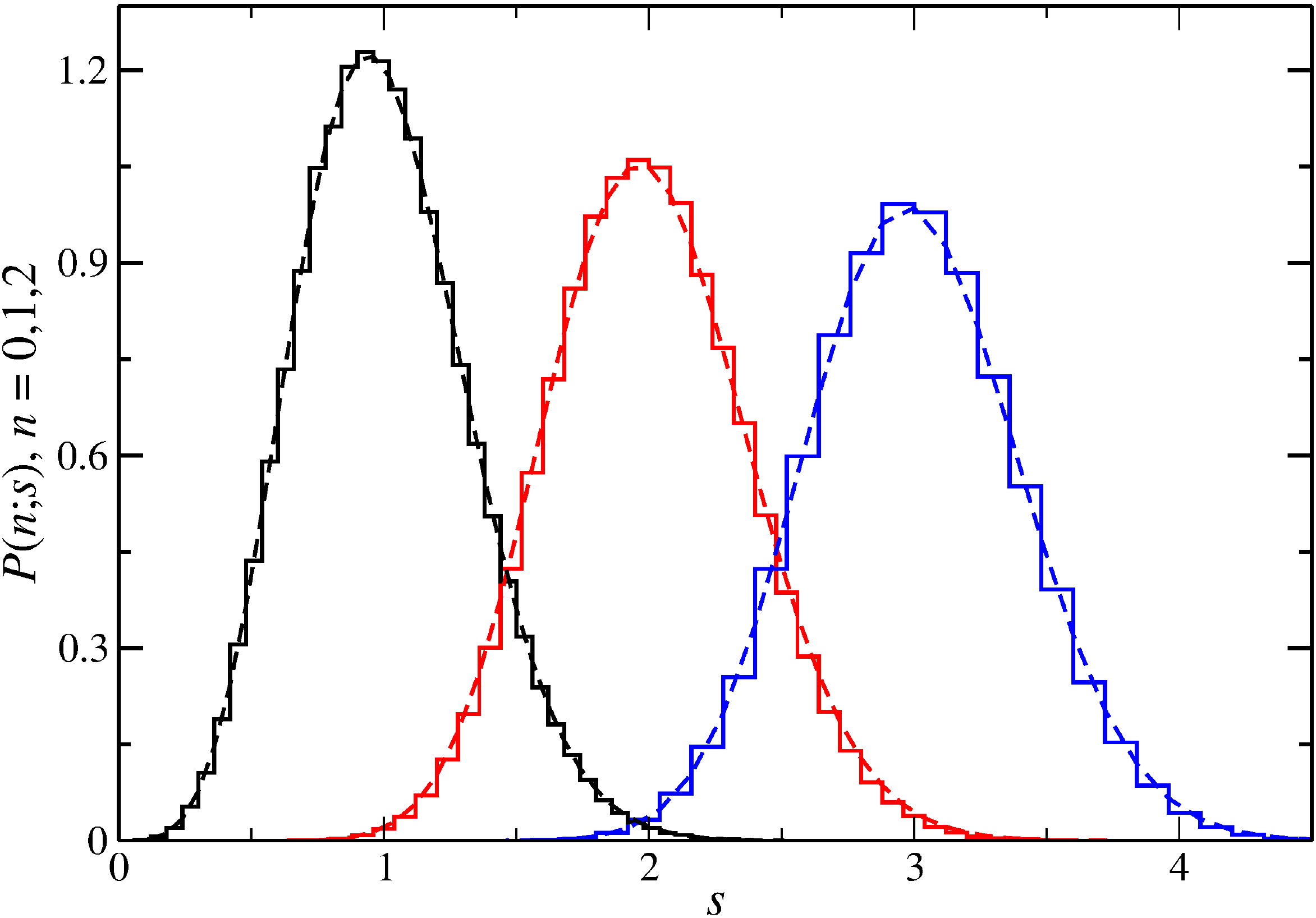}
\caption{Nearest-neighbor (black histogram), next-nearest neighbor (red histogram) and second-nearest neighbor (blue histogram) spacing distributions for the GSE. The dashed curves were obtained from a fit of $\tilde P(s)=\gamma s^\mu e^{-\chi s^2}$ to the histogram of corresponding color~\cite{Bohigas2006,Stoffregen1995}.}
\label{Fig2}
\end{figure}

For the GOE and GUE $P(1;s)$ is given in~\cite{Stoffregen1995}. We found out, that for the GSE the approximation of $P(n;s)$ by a Gaussian with variance $V^2(n)$ centered at $n+1$~\cite{Bohigas1999}, as done in the second sum in~\refeq{abst} is good for $n\geq 3$. Therefore, we chose $K=3$ and obtained $P(n;s)$ with $n=1,2$ by computing the ensemble averages of the normalized next and 2nd-next nearest-neighbor spacing distributions of 500 $500\times 500$ random matrices from the GSE, where the spacing $s$ was scaled to average spacing unity, and fitting $\tilde P(s)=\gamma s^\mu e^{-\chi s^2}$ to the resulting distributions, as illustrated in~\reffig{Fig2}. Furthermore, for $n\gtrsim 10$ the contributions to $p(s)$ are negligibly small, so that we chose $M=10$.   
For larger values of $n$ $P(n;s)$ is well approximated by a Gaussian with variance $V^2(n)$ centered at $n+1$~\cite{Bohigas1999}, thus yielding the second sum in~\refeq{abst}.
We considered spectra with up to $30\%$ missing levels and obtain good agreement with~\refeq{abst} for $M\lesssim 10$. Similarly, the number variance $\Sigma^2$, the rigidity $\Delta_3$ and the power spectrum~\cite{Molina2007} are deduced from those for complete spectra ($\Phi =1$),
\begin{equation}
\sigma^2(L)=(1-\Phi)L+\Phi^2\Sigma^2\left(\frac{L}{\Phi}\right),
\label{sigma2}
\end{equation}
\begin{equation}
\delta_3(L)=(1-\Phi)\frac{L}{15}+\Phi^2\Delta_3\left(\frac{L}{\Phi}\right).
\label{delta3}
\end{equation}
and
\begin{eqnarray}
s(\tilde\tau) &=&\nonumber
\frac{\Phi}{4\pi^2}\left[\frac{K\left(\Phi\tilde\tau\right)-1}{\tilde\tau^2}+\frac{K\left(\Phi\left(1-\tilde\tau\right)\right)-1}{(1-\tilde\tau)^2}\right]\\
&+& \frac{1}{4\sin^2(\pi\tilde\tau)} -\frac{\Phi^2}{12}.
\label{noise}
\end{eqnarray}
Here, $K(\tau)=1-b(\tau)$ denotes the spectral form factor where 
\be
b(\tau)=\int_{-\infty}^\infty Y_2(r)e^{-ir\tau}dr
\label{form}
\ee
is the Fourier transform of the two-point cluster function $Y_2(e_1,e_2)=Y_2(r=(e_1-e_2))$ which for the GUE equals~\cite{Mehta1990}
\be
Y_2(r)=\left(\frac{\sin[\pi r]}{\pi r}\right)^2
\ee
and for the GSE
\be
Y_2(r)=\left(\frac{\sin[2\pi r]}{2\pi r}\right)^2-\frac{d}{dr}\left(\frac{\sin[2\pi r]}{2\pi r}\right)\cdot\int_0^r\left(\frac{\sin[2\pi x]}{2\pi x}\right)
dx\, .
\ee
For the GUE $b(\tau)=1-\vert\tau\vert$ for $\vert\tau\vert\leq 1$ and zero otherwise, whereas for the GSE it vanishes for $\vert\tau\vert\geq 2$ and is given by 
\be
b(\tau)=1-\frac{1}{2}\vert\tau\vert+\frac{1}{4}\vert\tau\vert\ln\vert 1-\vert\tau\vert\vert
\label{form_gse}
\ee
otherwise. 

The two-point cluster function is obtained from the two-point correlation function by choosing in~\refeq{rn} $n=2$ and rescaling the eigenvalues $E_i$ to mean spacing unity, yielding  
\be
y_2(r)=Y_2\left(\frac{r}{\Phi}\right).
\ee
Using this feature of the two-point cluster function we computed from Eqs.~(\ref{sigma2}),~(\ref{delta3}) and~(\ref{noise}) RMT predictions for the variance $\Sigma^2(L)$, the rigidity $\Delta_3(L)$ and the power spectrum based on their relation to $Y_2(r)$, which for the latter is given in Eqs.~(\ref{noise}),~(\ref{form}) and (\ref{form_gse}), whereas
\be
\Sigma^2(L)=L-2\int_0^L(L-r)Y_2(r)dr\, ,
\ee
and 
\be
\Delta_3(L)=\frac{L}{15}-\frac{1}{15L^4}\int_0^L(L-r)^3\left(2L^2-9rL-3r^2\right)Y_2(r)dr\, .
\ee

\section{Fluctuation properties in the incomplete spectra of GUE and GSE quantum graphs\label{Test}}
\begin{figure}[!th]
\includegraphics[width=0.9\linewidth]{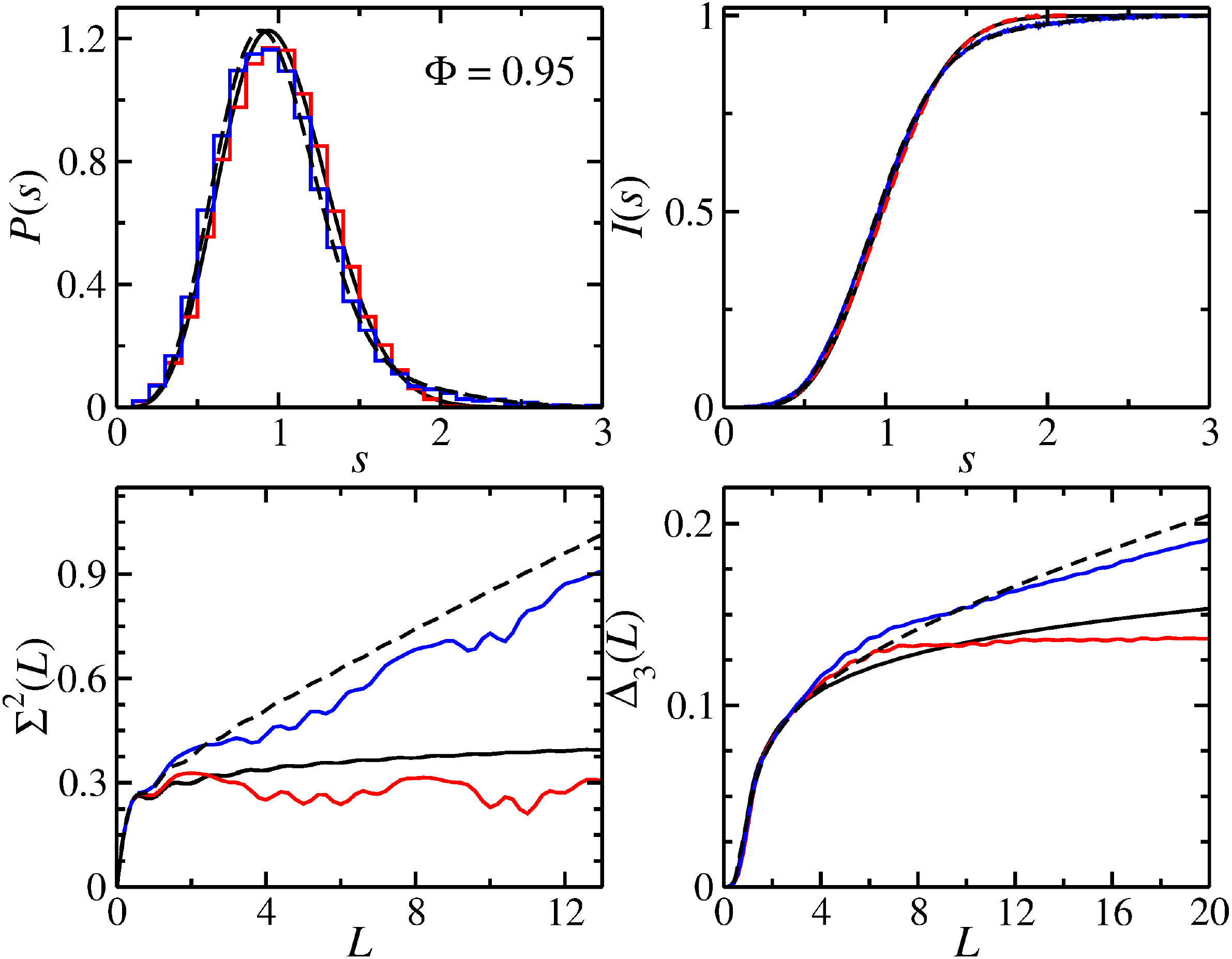}
\includegraphics[width=0.9\linewidth]{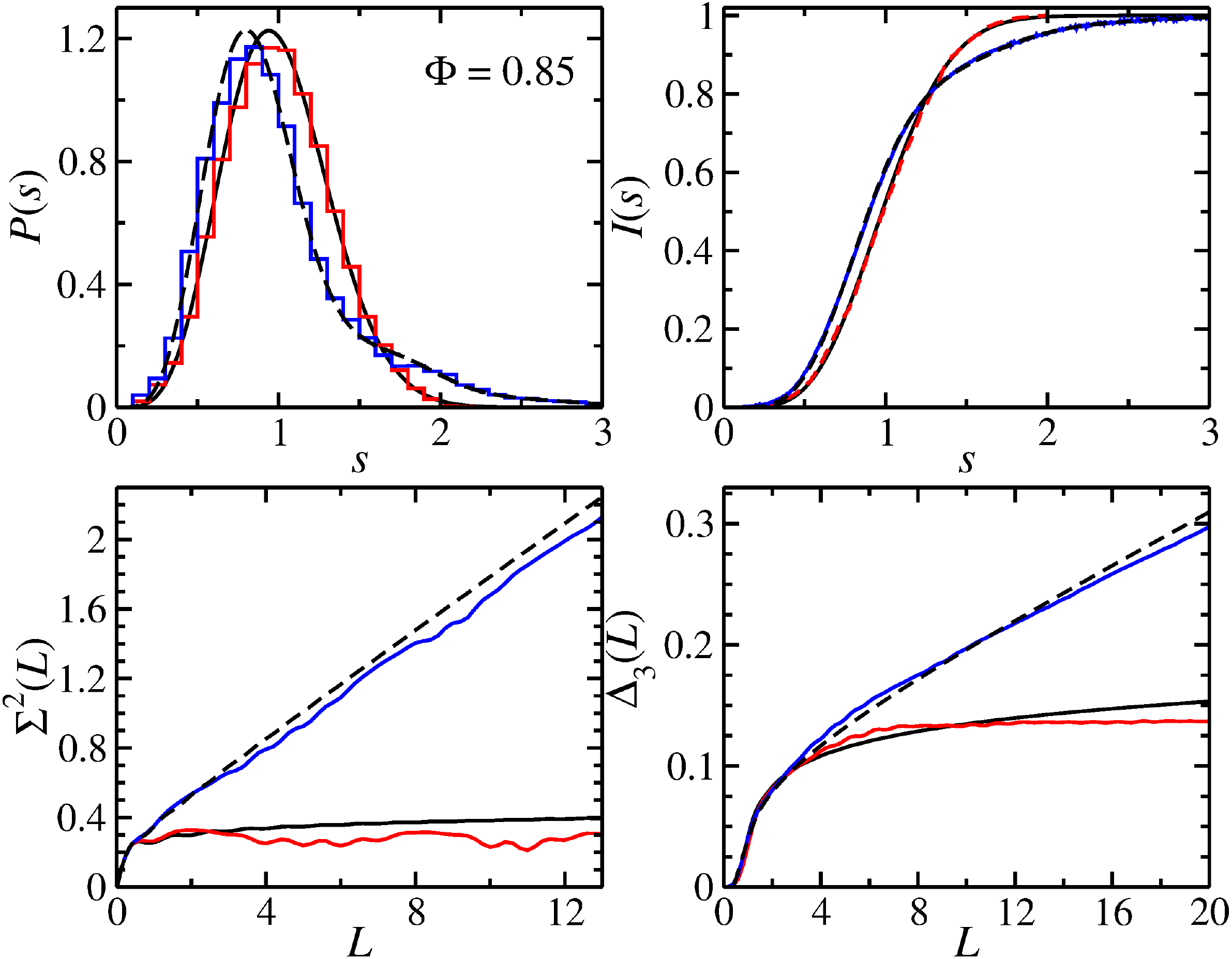}
\includegraphics[width=0.9\linewidth]{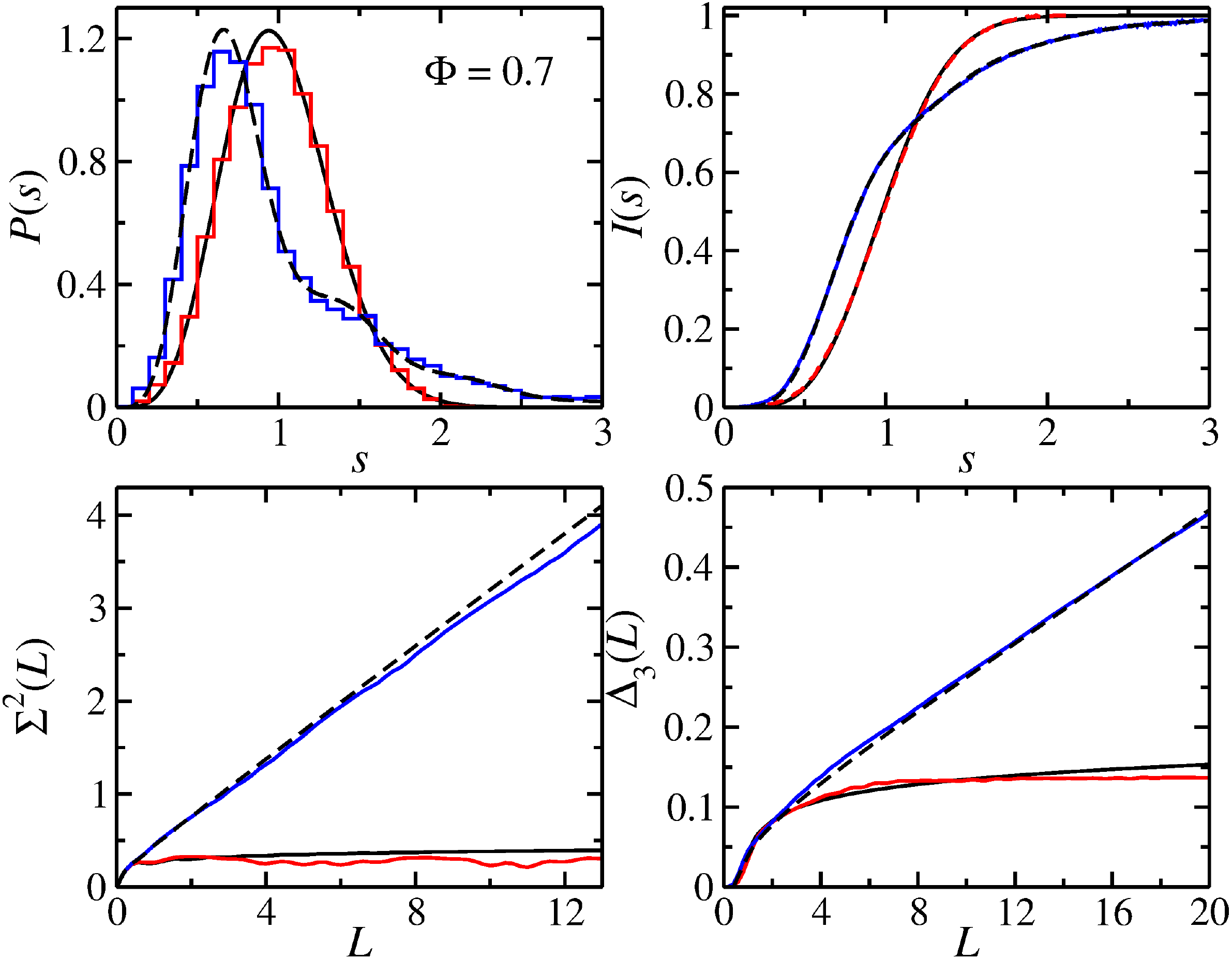}
	\caption{Comparison of the spectral properties of the numerically obtained eigenvalues of a quantum graph with symplectic symmetry with the corresponding GSE curves obtained from Eqs.~(\ref{abst})-(\ref{delta3}). The red (quantum graph) and black (RMT) curves show the results for complete eigenvalue sequences ($\Phi=1$). The blue (quantum graph) and dashed black (RMT) curves show the results after randomly extracting eigenvalues. The fraction of levels $\Phi$ taken into account is indicated in each panel.}
\label{Fig3}
\end{figure}

We first validated Eqs.~(\ref{abst})-(\ref{noise}) for incomplete spectra of quantum systems belonging to the symplectic universality class based on the numerically obtained level dynamics of GSE graphs. To ensure statistical independence, we considered an ensemble of 25 out of these 500 complete sequences comprising 1000 eigenvalues each and extracted up to 30$\%$ of them. The results are shown for $\Phi=0.95,\, 0.85,\, 0.7$ in~\reffig{Fig3} (blue curves). For comparison we also show the result for the complete sequence (red curves). They are compared to the GSE curves for complete spectra (full black lines) and to the curves obtained from Eqs.~(\ref{abst})-(\ref{delta3}) (dashed black lines). For the nearest-neighbor spacing distribution the agreement between numerics and RMT predictions is good for complete and incomplete spectra. However, for the number variance $\Sigma^2(L)$ and rigidity $\Delta_3(L)$ the agreement is only good below $L\simeq 2-3$~\cite{Dietz2017,Lu2020}, whereas for $L\gtrsim 2-3$ the curves lie above the RMT predictions for both complete and incomplete spectra. These discrepancies were shown to occur due to the presence of periodic orbits that are confined to individual bonds by backscattering at the vertices bordering them or to loops, that is, to a part of the quantum graph. They correspond to wave functions in the associated quantum graph that are localized on the individual bonds or on loops within the graph, and thus do not sense the chaoticity of the underlying classical dynamics resulting from scattering at all vertices~\cite{Dietz2017,Lu2020}. In distinction to the bouncing-ball orbits in the stadium billiard~\cite{McDonald1979,Graef1992,Sieber1993} the nonuniversal contributions of these periodic orbits~\cite{Kottos1999} can only partly be removed since their number is large~\cite{Lu2020}. 

Deviations due to nonuniversal contributions are similar in size for the complete and incomplete spectra, and their effect on $\Delta_3(L)$ is larger than on $\Sigma^2(L)$. Note that the former is given as an integral over the latter. Integration corresponds to a smoothing of the oscillatory features in $\Sigma^2(L)$ and we may conclude that thereby deviations originating from the nonuniversal contributions are enhanced. Yet, the effect of the nonuniversal contributions on the spectral properties is nonnegligible only for spacings $s\gtrsim 1$ or distances $L\gtrsim 2-3$, and does not distort the characteristic behavior, i.e., the degree of level repulsion, which manifests itself in the shapes of $P(s)$, $\Sigma^2(L)$ and $\Delta_3(L)$ for small $s$ and $L$, respectively, and depends on the universality class of the underlying quantum systems. Thus, the unambiguous determination of the universality class of a quantum graph from the statistical measures is possible despite the nonuniversal features~\cite{Gnutzmann2004,Pluhar2014} for complete and incomplete sequences. Beyond a certain value of $L$ $\Sigma^2(L)$ and $\Delta_3(L)$ saturate below the RMT prediction as expected for the long-range correlations of pairs of eigenvalues of any generic quantum system if their distance is beyond a certain number $L$ of mean spacings, which is inversely proportional to the length of the shortest periodic orbit~\cite{Berry1985,Sieber1993} or for spectra of short length.

\begin{figure}[!th]
\includegraphics[width=0.9\linewidth]{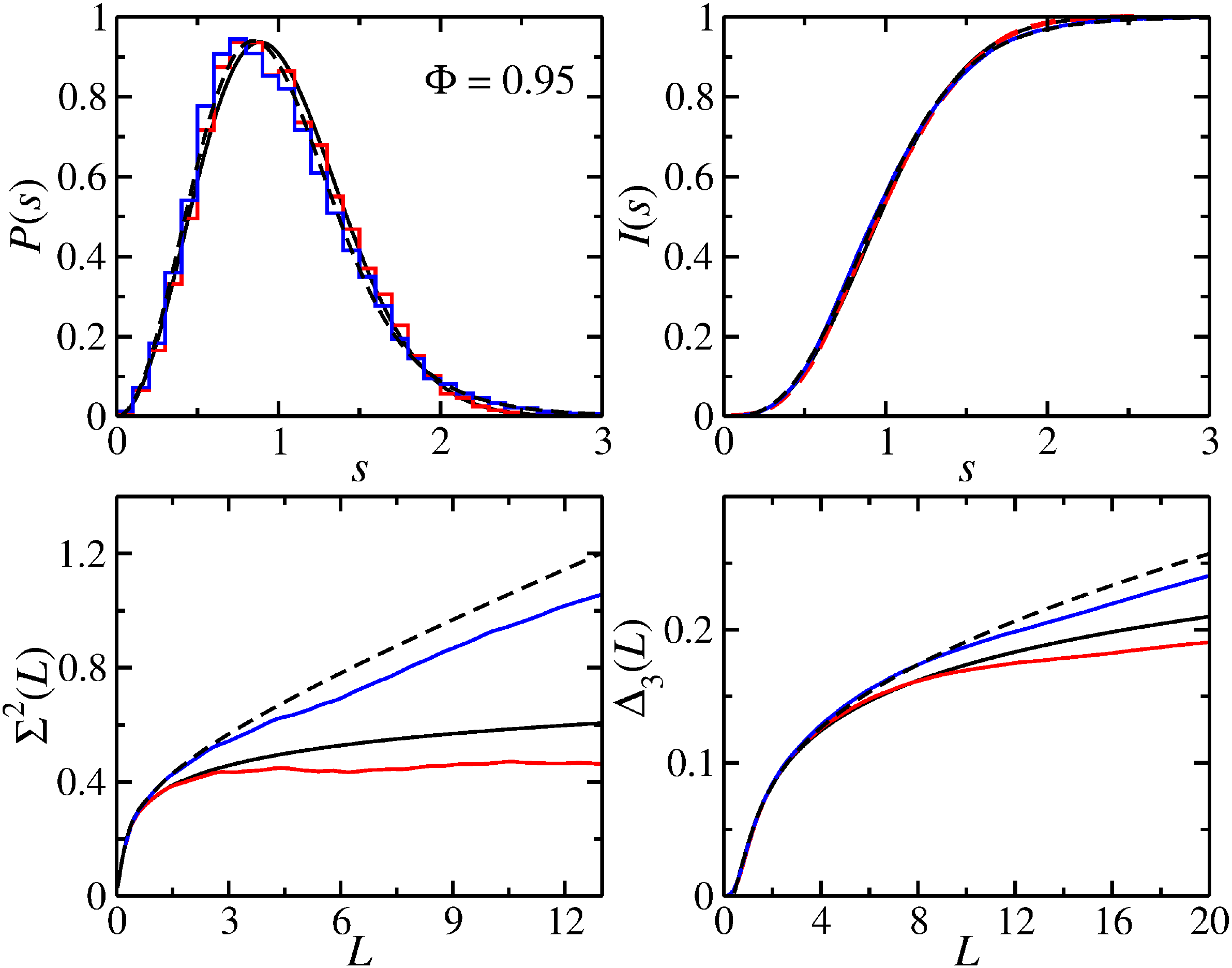}
\includegraphics[width=0.9\linewidth]{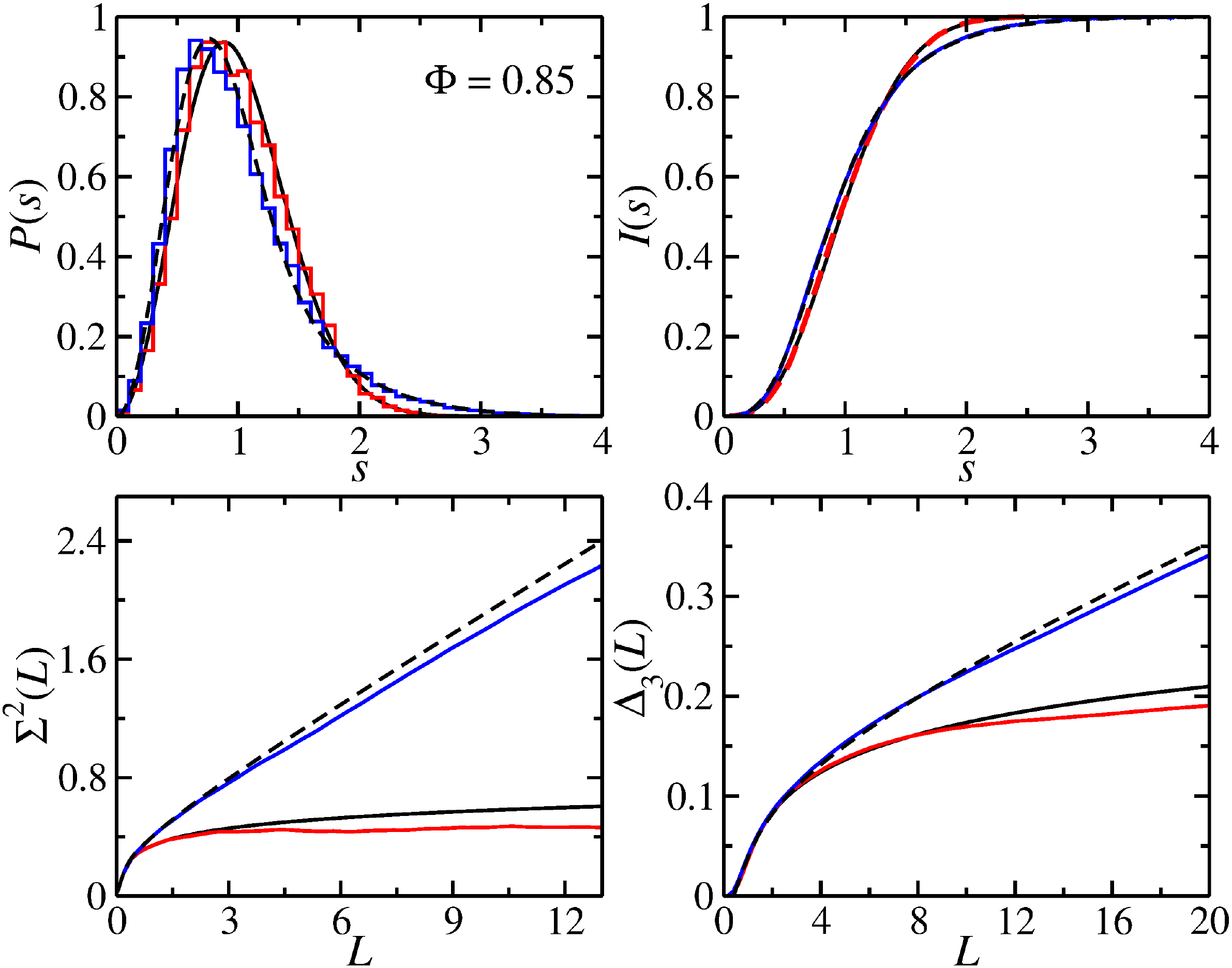}
\includegraphics[width=0.9\linewidth]{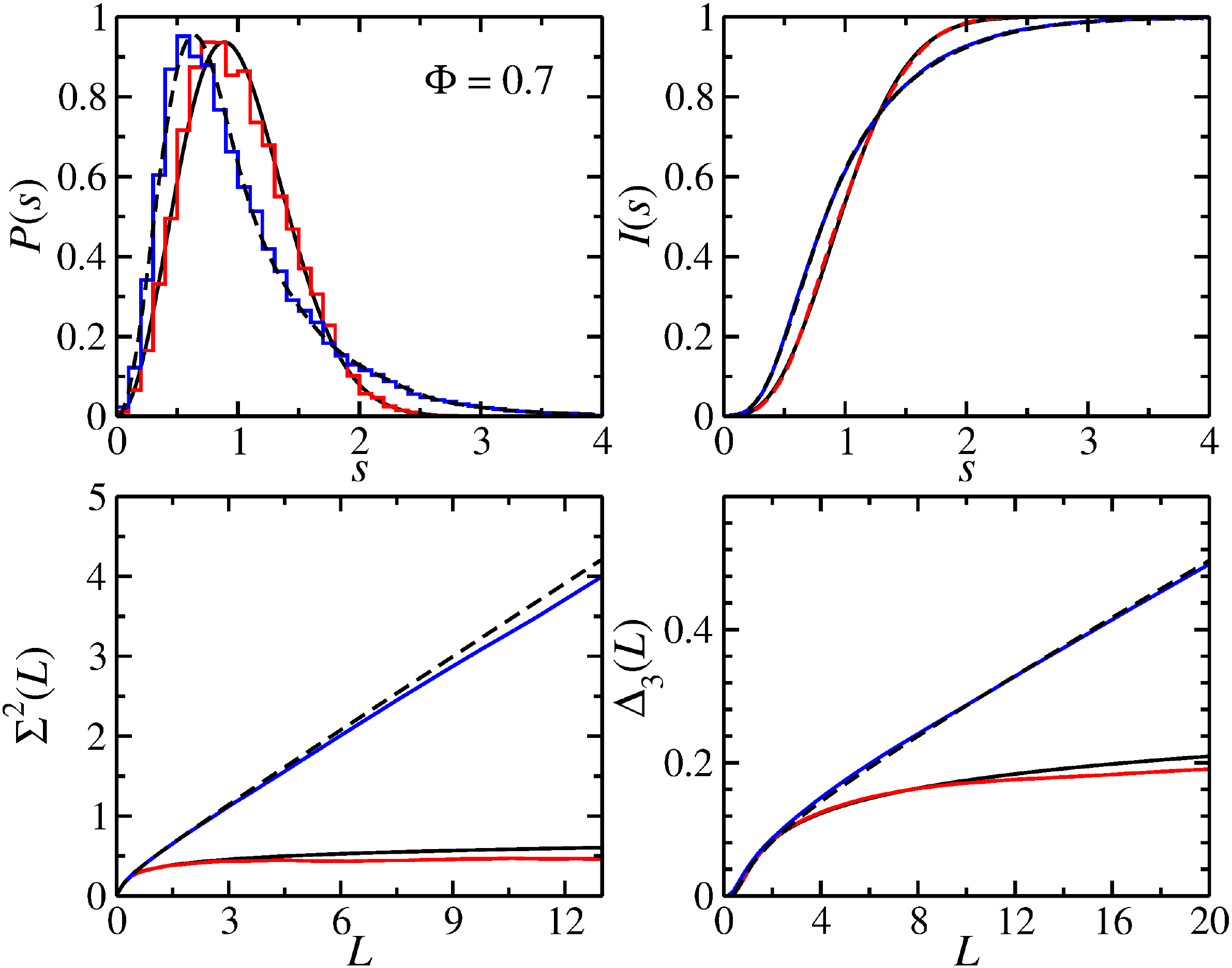}
	\caption{Same as~\reffig{Fig3} for a GUE quantum graph with unitary symmetry. The numerical results are compared to those deduced from~\refeq{noise}.}
\label{Fig4}
\end{figure}

Since the asymptotic behavior of the power spectrum $\tilde S(\tilde\tau)$ for small values of $\tilde\tau$ of a quantum graph with chaotic classical counterpart does not depend on the universality class, the fraction of missing levels can be unambiguously determined from it, also from $P(s)$ and $I(s)$ since the effect of missing levels on them is particularly large for GSE quantum graphs. To illustrate this we show in~\reffig{Fig4} corresponding curves for GUE graphs, which are based on complete eigenvalue spectra computed in~\cite{Lu2020}. Red and blue curves correspond to the numerical results for complete and incomplete spectra, respectively, black full and dashed lines to the corresponding results for the GUE. The effect of missing levels on GUE and GSE graphs is comparable for the long-range correlations, whereas that on the nearest-neighbor spacing distribution is considerably larger for GSE graphs. Deviations due to the nonuniversal contributions of periodic orbits confined to a fraction of the GUE graph are again visible in the long-range correlations and comparable for complete and incomplete spectra. 

\begin{figure}[!th]
\includegraphics[width=0.9\linewidth]{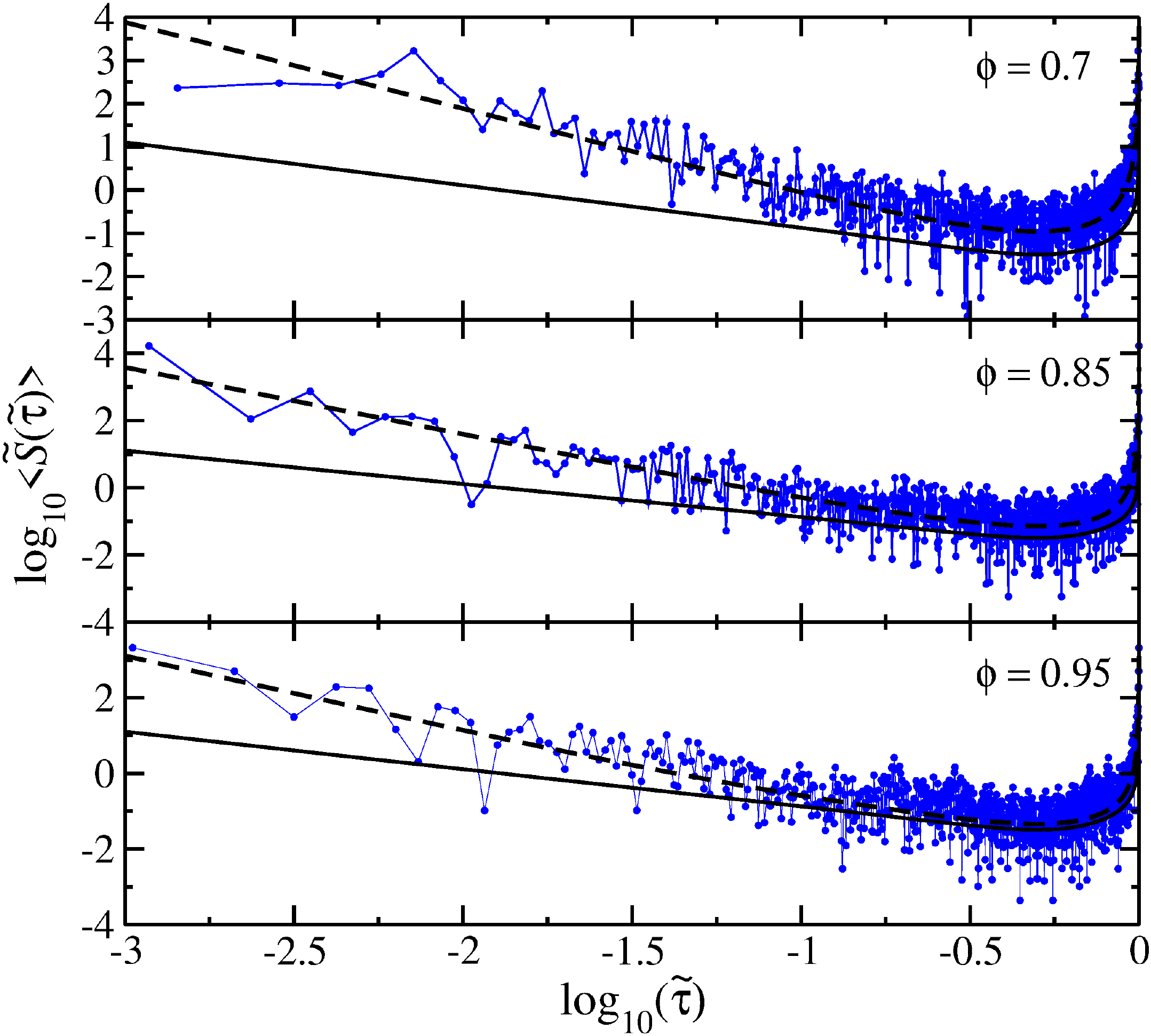}
\caption{Power spectrum of the quantum graph (blue dots and dashed lines) for the same sets of eigenvalues as in~\reffig{Fig3} and the corresponding GSE curves. The fraction of eigenvalues taken into account is indicated in each panel.}
\label{Fig5}
\end{figure}

In~\reffig{Fig5} we compare the numerical results for the power spectrum (blue) to the RMT predictions~\refeq{noise} (black dashed lines). To illustrate the effect of missing levels we also show the GSE prediction for complete spectra (black full line). The agreement is very good, which implies that the effect from nongeneric contributions is diminished. We may conclude that $s(\tilde\tau)$ is particularly suited for the determination of the fraction of missing levels, since its asymptotic bevahior does not depend on the universality class, thus confirming the supposition of~\cite{Bialous2016} and then $\Sigma^2(L)$ and $P(s)$ may be used to obtain or confirm the universality class from their behavior at small $L\lesssim 2-3$ and $s\lesssim 1$, respectively.

\begin{figure}[!th]
\includegraphics[width=0.9\linewidth]{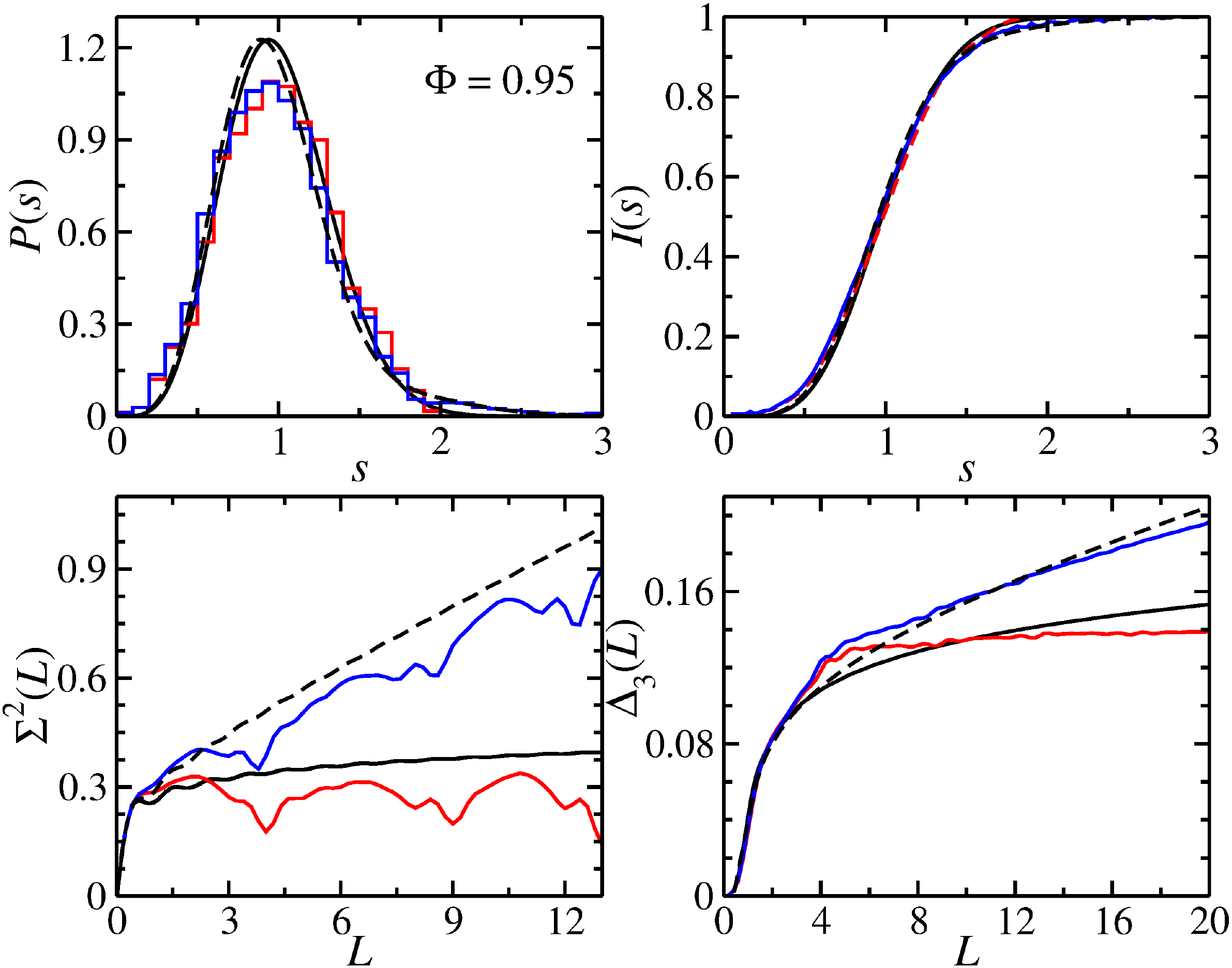}
\includegraphics[width=0.9\linewidth]{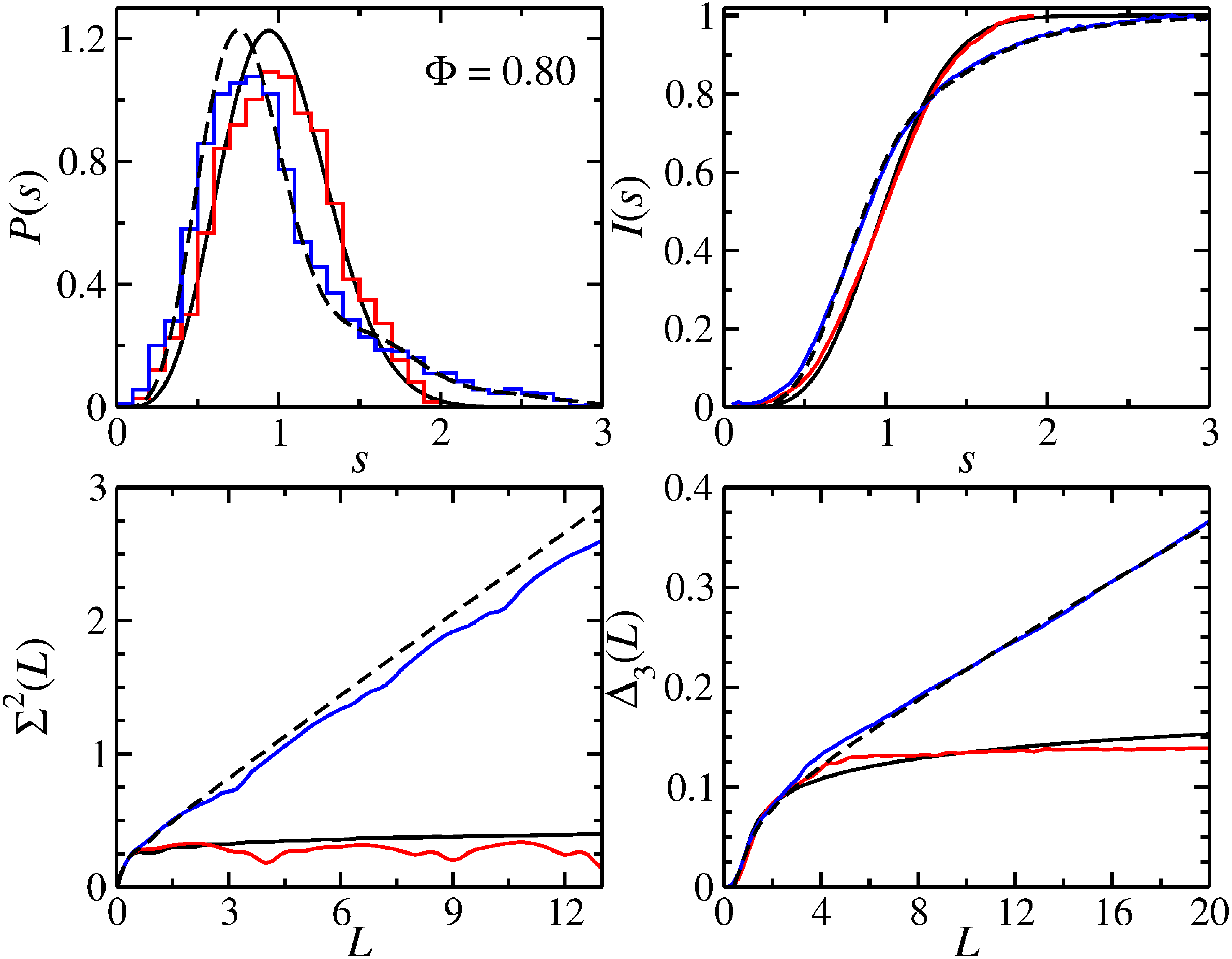}
\includegraphics[width=0.9\linewidth]{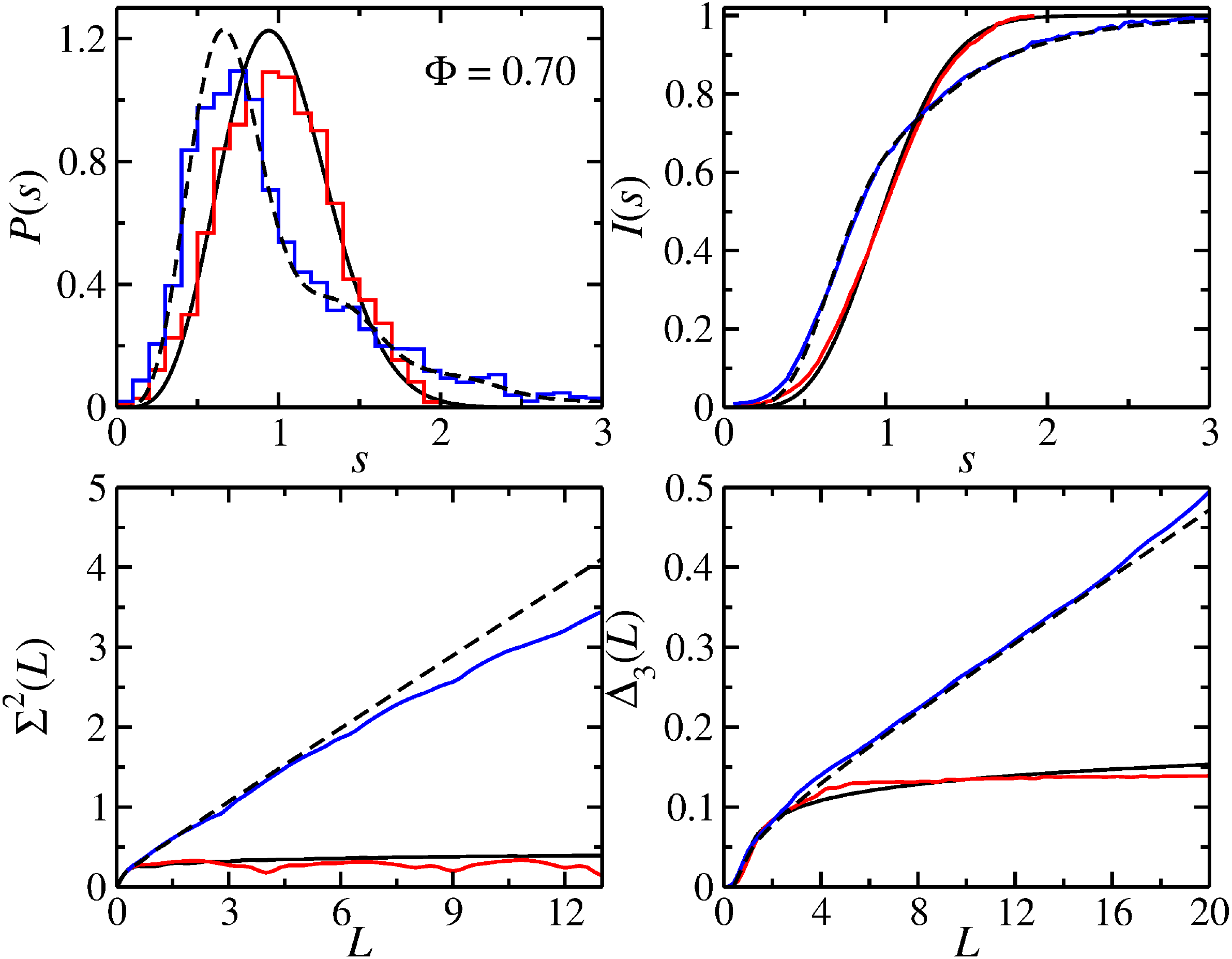}
\caption{Same as~\reffig{Fig3} for the experimentally determined eigenfrequencies of microwave networks with symplectic symmetry presented in Ref.~\cite{Lu2020}. Here, eigenfrequencies were extracted randomly from the complete sequences. The fraction of eigenfrequencies is indicated in the panels.}
\label{Fig6}
\end{figure}

\begin{figure}[!th]
\includegraphics[width=0.9\linewidth]{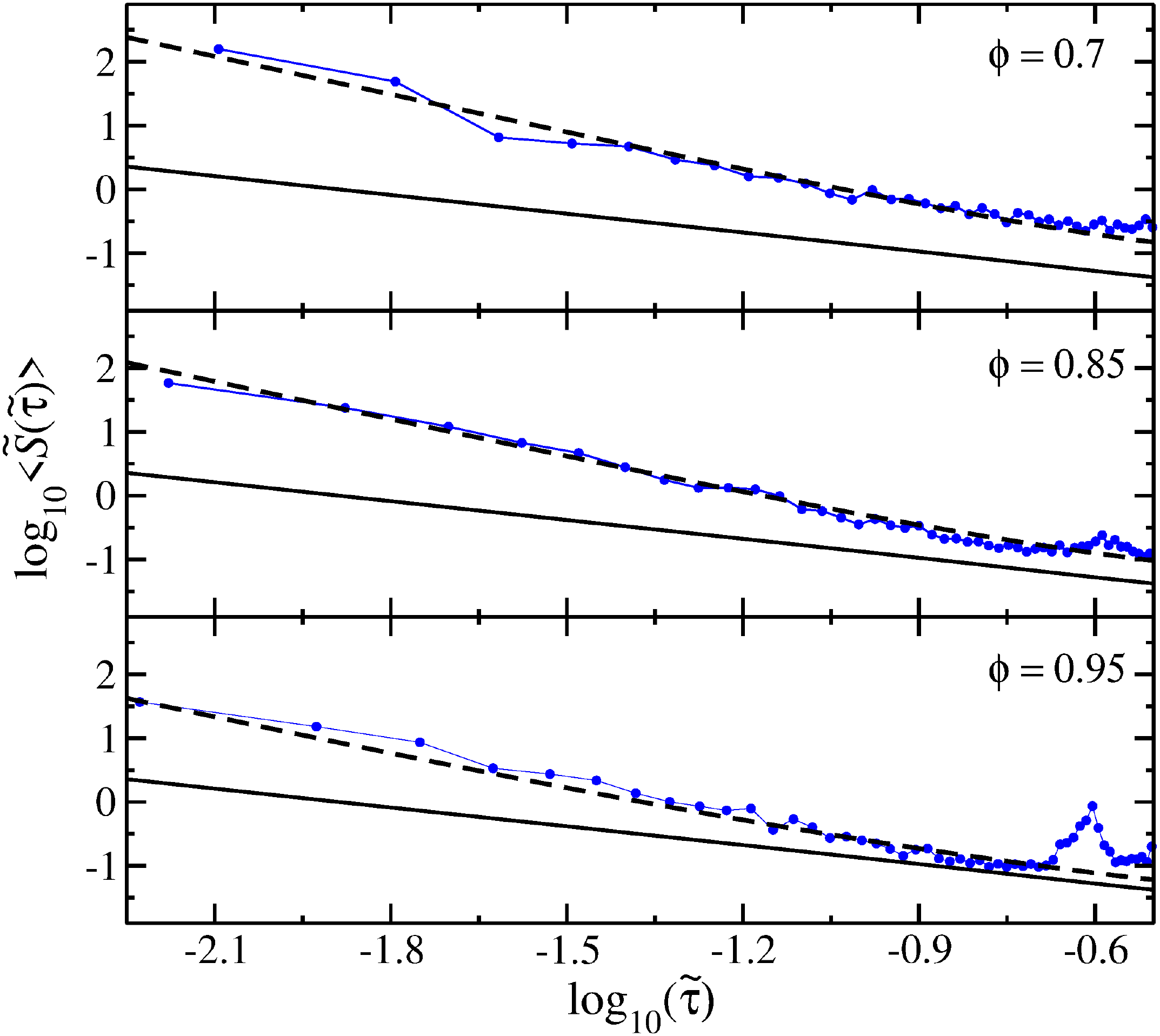}
\caption{Same as~\reffig{Fig5} for the power spectrum.}
\label{Fig7}
\end{figure}

In order to validate the RMT predictions Eqs.~(\ref{abst})-(\ref{noise}) experimentally, we performed a similar analysis for the 43 complete sequences of 178 eigenfrequencies which were determined in the experiments described in Ref.~\cite{Lu2020}, that is, we randomly extracted from each spectrum up to 30$\%$ of the eigenfrequencies and analysed their spectral properties. Results are shown in Figs.~\ref{Fig6} and~\ref{Fig7}. To illustrate the effect of missing levels (blue curves) we also show the result for the complete spectra (red curves). Again deviations due to nonuniversal contributions are especially visible in the number variance $\Sigma^2(L)$ and rigidity $\Delta_3(L)$ and are comparable in size for the complete and incomplete spectra. Deviations between the experimental nearest-neighbor spacing distribution and the RMT prediction~\refeq{abst} for small spacings are attributed to experimental inaccuracy. Except for these discrepancies the agreement between the experimentally obtained curves and the RMT predictions is as good as for the numerically obtained ones, thus validating the RMT predictions Eqs.~(\ref{abst})-(\ref{delta3}). However, discrepancies are clearly visible in~\reffig{Fig7} for the power spectrum around $\log_{10}(\tilde\tau)\simeq -0.6$. They, actually, are largest for the spectra with smallest number of extracted eigenfrequencies, i.e., largest value of $\Phi$ and therefore may be attributed to the nongeneric contributions. 
\begin{figure}[!th]
\includegraphics[width=0.9\linewidth]{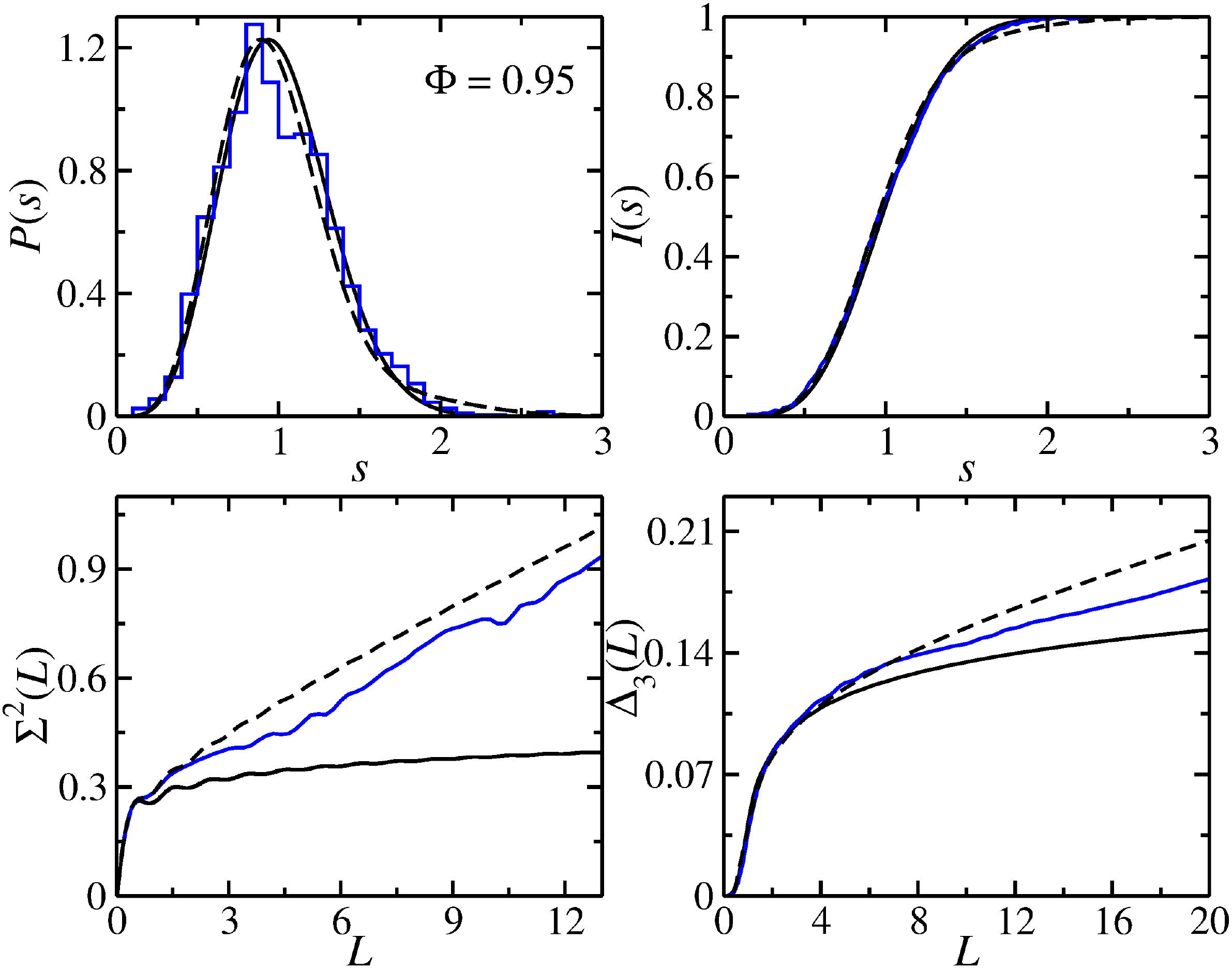}
	\caption{Spectral properties obtained from an ensemble average over the statistical measures deduced from eigenfrequency sequences obtained from measurements with 10 different realizations of microwave networks with symplectic symmetry. In all sequences approximately 5 $\%$ of the eigenfrequencies are missing. The black solid curves show the RMT predictions for complete eigenvalue sequences, the dashed black curves show the corresponing results the incomplete sequence with $\Phi=0.95$.}
\label{Fig8}
\end{figure}

\begin{figure}[!th]
\includegraphics[width=0.9\linewidth]{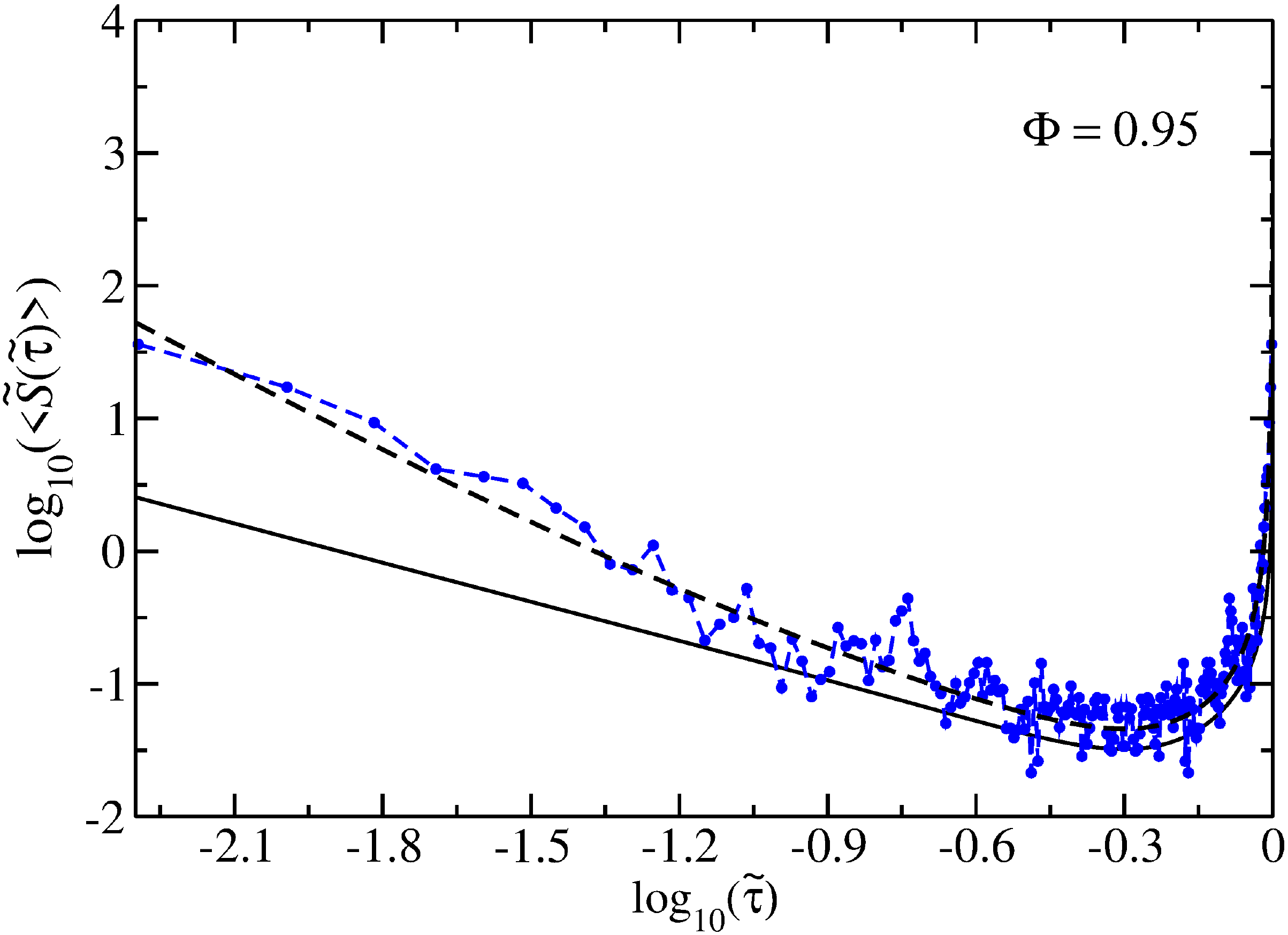}
\caption{Same as~\reffig{Fig8} for the power spectrum.}
\label{Fig9}
\end{figure}
In further studies we used incomplete spectra that were obtained from 10 independent measurements with microwave networks simulating distinct GSE quantum graphs. According to Weyl's formula~\refeq{Weyl} about $5\%$ of the eigenfrequencies are missing in each spectrum. Note, that the nearest-neighbor spacing distribution increases around $s\simeq 0$ as $P(s)\propto s^4$ for the GSE, implying that the probability that the spacing between neighboring minima in the reflection amplitude $S_{11}(\nu)$ is much less than the average spacing is small for GSE graphs and thus should ease the identification of eigenfrequencies of GSE microwave networks. This indeed is the case when comparing the efforts needed to determine the eigenfrequencies to those for GOE and GUE graphs. Still, due to absorption in the coaxial cables we were not able to determine complete sequences for individual graphs, that is, in order to identify all eigenfrequencies we would need to proceed as in~\cite{Lu2020} which is time consuming. For this reason the development of theoretical results for the case of incomplete spectra is indispensable. Indeed, the agreement between the statistical measures obtained by averaging over the ensemble of 10 independent GSE quantum graphs (blue curves in Figs.~\ref{Fig8} and~\ref{Fig9}) with RMT predictions Eqs.~(\ref{abst})-(\ref{noise}) for $\Phi=0.95$ (black dashed lines) is as good as in Figs.~\ref{Fig3}-\ref{Fig7}.  
\begin{figure}[!th]
\includegraphics[width=0.9\linewidth]{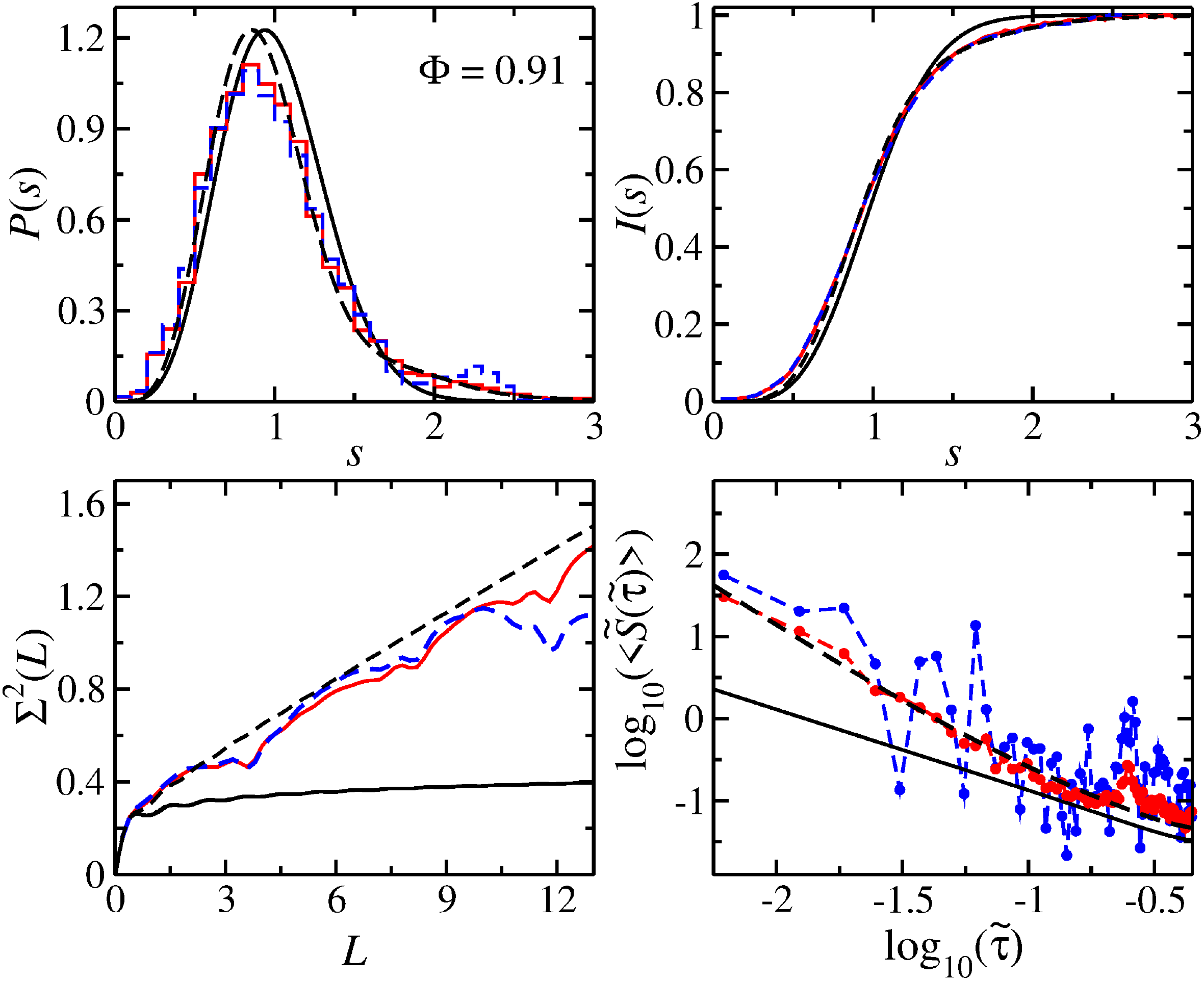}
	\caption{Fluctuation properties in the experimentally determined eigenfrequency sequences of the microwave networks with symplectic symmetry presented in Ref.~\cite{Lu2020} after extraction of nonuniversal contributions (blue dashed lines and dots) (see main text) and randomly extracting (red curves) a fraction $\Phi =0.91$ of the levels.}
\label{Fig10}
\end{figure}

In Ref.~\cite{Lu2020} nongeneric contributions of modes that are localized on individual bonds could be extracted by exploiting the fact that the associated eigenfrequencies do not change, when the length of another bond is varied to generate a parametric eigenfrequency sequence, or a local perturbation is induced when the length of the bond on which the wave function is localized is varied. Accordingly, the eigenfrequencies corresponding to these localized wave functions do not change with the parameter and thus they could be identified and extracted. For a fixed parameter the spectral properties will be effected because of incompleteness of the level sequence, thus demanding an analysis as proposed in this paper. Note, that it is difficult to identify eigenfrequencies with wave functions that are localized on a few bonds, and thus it is impossible to remove all nongeneric contributions. For the GSE graph studied in~\cite{Lu2020} about $9\%$ of the eigenfrequencies correspond to such parameter-independent states, so that their extraction leads to an incomplete spectrum with $\Phi=0.91$. In~\reffig{Fig10} we compare the resulting curves (blue) to the corresponding RMT predictions. Discrepancies are clearly visible in the nearest-neighbor spacing distribution for large spacings. These may be attributed to the fact that the RMT predictions are applicable to incomplete spectra where levels are randomly missing, whereas eigenfrequencies were extracted systematically in the sense that the same eigenfrequencies were removed from each sequence of the ensemble leading to gaps in the level dynamics. To see the difference between both procedures of removing levels from a complete spectrum we show as red curve the result for randomly extracted eigenfrequencies. The agreement with the RMT predictions is similar and even better for the latter case for larger spacings $s$ and $L$, respectively. This implies that we do not obtain an improvement for the agreement between the experimental curves and RMT predictions after partly removing nongeneric contributions, which is in contrast to the findings for parametric spectral properties~\cite{Lu2020}, a reason being that the latter do not rely on completeness of the spectra. Nevertheless, for small values of $s$ and $L$ agreement with the RMT predictions allows to specify the universality class of the quantum graph and fraction of missing levels.  

\section{Conclusions\label{Concl}}
We extended the missing-level statistics approach introduced in Ref.~\cite{Bohigas2006} to derive statistical measures for the fluctuation properties in incomplete spectra of quantum graphs belonging to the symplectic universality class. We validated them based on ensembles of eigenfrequency sequences obtained numerically for several GSE quantum graphs and experimentally for microwave networks. The data sets were attained recently for the investigation of parametric properties in GSE and GUE graphs~\cite{Lu2020}. The derivation of the RMT predictions is applicable to generic quantum systems and relies on the assumption that eigenfrequencies are missing randomly, which might be a drawback when comparing to experimental data. In the experiments with microwave networks eigenfrequencies may not be detected if the spacing between them and adjacent ones is too small, or if the microwave intensity vanishes at the position of an antenna. Yet, for GSE graphs, the probability of close lying eigenfrequencies is very small and thus randomness may be complied with by considering ensembles of statistically independent spectra. However, the spectra of quantum graphs comprise nongeneric contributions originating from eigenstates with wave functions that are localized on individual bonds or on a fraction of the quantum graph that can only partly be removed~\cite{Lu2020}. Extraction of these nongeneric contributions corresponds to removing the same eigenfrequencies from each of the sequences forming the level dynamics and thus to a systematic missing of levels. Still, after extracting part of the nongeneric contributions the short- and long-range correlations between eigenfrequencies are in accordance with RMT predictions for moderate spacings between them and similar to those obtained by randomly removing eigenfrequencies. Generally, despite the presence of nongeneric contributions we find good agreement between the spectral properties of the experimental microwave networks and numerical quantum graphs and the RMT predictions for the nearest-neighbor spacing distributions and in the long-range correlations for values of $L$ corresponding to 2-3 mean spacings, and thus demonstrate the applicability of the missing-level statistics approach introduced in Ref.~\cite{Bohigas2006}. Since the power spectrum~\refeq{PowerS} asymptotically exhibits a power-law behavior which does not depend on the universality class, this measure is suitable for the determination or confirmation of the fraction of missing levels~\cite{Bialous2016} obtained based on Weyl's law~\refeq{Weyl}. Then, the universality class may be obtained by comparing the nearest-neighbor spacing distribution $P(s)$ for $s\lesssim 1$ and the number variance $\Sigma^2(L)$ for $L\lesssim 2-3$ with the RMT predictions for the GOE, GUE and GSE, respectively, since in these ranges of $s$ and $L$, contributions from the nonuniversal features are negligible~\cite{Gnutzmann2004,Pluhar2014}. These findings are of particular importance for the still ongoing experimental studies of quantum systems with symplectic symmetry where one has to cope with the problem that generally the identification of complete sequences of levels is impossible. On the other hand, the analytical results presented in this article can be used to unambiguously verify the chaoticity of the classical dynamics and determine the universality class based on the complete sequence of a quantum system by extracting a certain fraction of levels and comparing with these prediction.  

\section{Acknowledgement}
This work was supported by the NNSF of China under Grant Nos. 11775100, 11961131009 and 12047501. 

\bibliography{ref_spaghetti,ref_triv,References}
\end{document}